1  NATURE-BASED SOLUTIONS IN THE URBAN CONTEXT: TERMINOLOGY, CLASSIFICATION AND SCORING FOR

2  URBAN CHALLENGES AND ECOSYSTEM SERVICES.

3  J.A.C. Castellar[a,b]*, L.A. Popartan[c], J. Pueyo-Ros[a,b], N. Atanasova[d], G. Langergraber[e], I. Säumel[f], L.

4  Corominas[a,b],  J. Comas[a,c], V. Acuña [a,b]

5  [a] Catalan Institute for Water Research (ICRA). Carrer Emili Grahit 101, 17003 Girona, Spain.

6  [b] University of Girona. Plaça de Sant Domènec 3, 17004 Girona, Spain.

7  [c] LEQUIA. Institute of the Environment, University of Girona Campus Montilivi, carrer Aurèlia Capmany, 69 E-17003 Girona.

8  Catalonia. Spain

9  [d] University of Ljubljana, Faculty of Civil and Geodetic Engineering. Jamova 2, 1000 Ljubljana, Slovenia

10 [e] Institute of Sanitary Engineering and Water Pollution Control, University of Natural Resources and Life Sciences, Vienna (BOKU),

11 Muthgasse 18, A-1190 Vienna, Austria

12 [f] Integrative Research Institute on Transformations of Human-Environment Systems (IRI THESys), Humboldt Universität zu Berlin,

13 Unter den Linden 6, D-10099 Berlin, Germany.

15 *Corresponding author: J.A.C Castellar. Email: jcastellar@icra.cat. Address:  Edifici H2O, Carrer Emili Grahit, 101,

16 17003 Girona

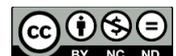

## Graphical abstract

**Towards a common understanding**

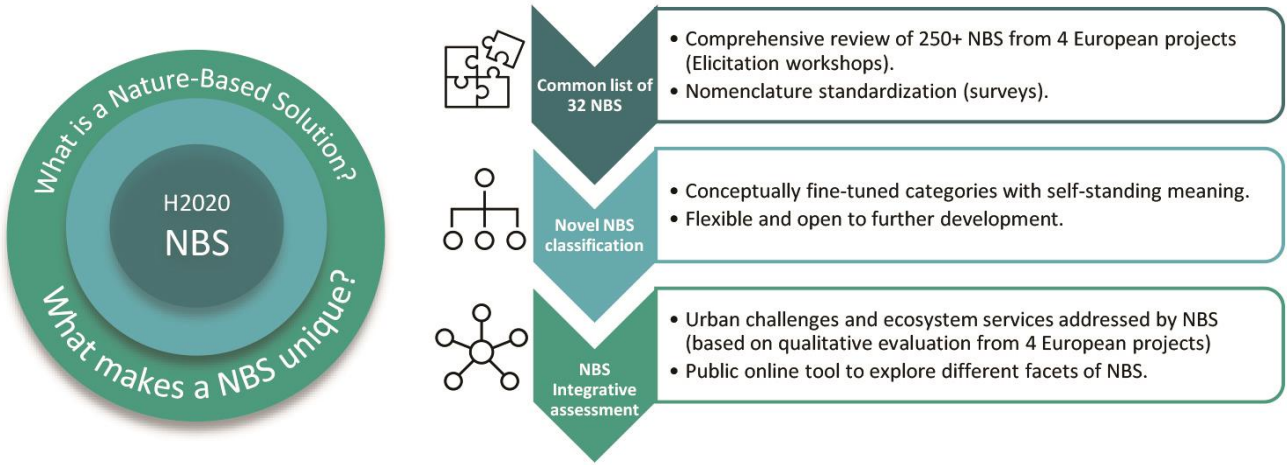

## Abstract

 The concept of Nature-Based Solutions (NBS) has emerged to foster sustainable development by

 transversally addressing social, economic, and environmental urban challenges. However, there is still a

 considerable lack of agreement on the conceptualization of NBS, especially concerning typologies,

 nomenclature, and performance assessments in terms of ecosystem services (ES) and urban challenges (UC).

 Therefore, this article consolidates the knowledge from 4 European projects to set a path for a common

 understanding of NBS and thus, facilitate their mainstreaming. To do so, firstly, we performed elicitation

 workshops to develop an integrative list of NBS, based on the identification of overlaps among NBS from

 different projects. The terminologies were formalized via web-based surveys. Secondly, the NBS were

 clustered, following a conceptual hierarchical classification. Thirdly, we developed an integrative assessment

 of NBS performance (ES and UC) based on the qualitative evaluations from each project. Afterwards, we run

 a PCA and calculated the evenness index to explore patterns among NBS. The main conceptual advancement

 resides in providing a list of 32 NBS and putting forward two novel NBS categories: NBS units (NBSu) that are

 stand-alone green technologies or green urban spaces, which can be combined with other solutions (nature-based or not); NBS interventions (NBSi) that refer to the act of intervening in existing ecosystems and in NBSu, by applying techniques to support natural processes. The statistical analysis suggests that NBSu are more versatile than NBSi in terms of UC and ES. Moreover, the results of the integrative assessment of NBS performance suggest a greater agreement concerning the role of NBS in addressing environmental UC, cultural and regulating ES than regarding socio-economic UC and supporting and provision ES. Finally, the 'green factor' and the replication of non-intensive practices occurring in nature seem to be key criteria for practitioners to identify a particular solution as an NBS.

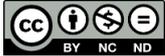

# 1. Introduction

More sustainable and resilient cities have been increasingly associated with a gradual (re) integration of nature into urban areas, overturning the dominance of grey infrastructure in city landscapes (Lafortezza et al., 2018). In this context, the concept of Nature-Based Solutions (NBS) has gained considerable popularity as it is deemed to facilitate the transition towards greener, more resilient and socially inclusive cities. The European Commission (EC) definition of NBS takes a broad perspective, referring to the three pillars of sustainability - economic, environmental and social (European Commission, 2015), while the International Union for Conservation of Nature (IUCN) definition (Cohen-Shacham et al., 2016) emphasizes actions for conservation and restoration (Albert et al., 2019; Carsten et al., 2017). Given the relative novelty of the concept and its accelerated uptake, it is no surprise that more definitions have arisen in the past years (Maes and Jacobs, 2017; van der Jagt et al., 2017; Short et al., 2019; Albert et al., 2019; for a review see Sarabi et al., 2019 and Carsten et al., 2017). Recently, the COST Action "Circular City" proposed a definition bringing a conceptual added value: It transfers the NBS concept into urban areas, putting a special emphasis on resource circularity (Langergraber et al., 2020). Nevertheless, this definitional 'wealth' and a lack of agreement on specific characteristics, unique to NBS (Dorst et al., 2019; Carsten et al., 2017), has led to a confusion within the NBS community concerning what is and what is not an NBS. The framing of the NBS concept does not always distinguish it from other well-established environmental concepts - such as ecological engineering, green infrastructure, urban green (and blue) spaces, and ecosystem-based adaptation, with which NBS shares key elements (Almenar et al., 2021; Sarabi et al., 2019). While the IUCN states that NBS is an umbrella concept that encompasses these similar approaches (Cohen-Shacham et al., 2016), others appeal to the need for a separate definition of NBS (Sarabi et al., 2019; Dorst et al., 2019; Carsten et al., 2017). The vagueness of NBS semantics is reflected in the way that different sets of NBS are classified and assessed in terms of Urban Challenges (UC) and Ecosystem Services (ES), especially in the framework of projects funded under the European Horizon 2020 (H2020) program. Moreover, there is a lack of agreement concerning the terminology used to refer to specific NBS. This means that one NBS may be referred under several different names. For example, while some authors may use the term "green façade with climbing plants" (URBANGREENUP, 2018), others might describe the same NBS as a "climber green wall" (NATURE4CITIES, 2020; Somarakis et al., 2019), and yet others may refer to this as "ground-based greening" (UNALAB, 2019) or green façade (Manso and Castro-Gomes, 2015). Therefore, to address this gap, the aim of the paper is to take a step towards a common understanding of the NBS typologies, terminologies, classification and evaluation. To do so, in this paper we integrate the vast knowledge across NBS reported in recently completed or ongoing H2020 European projects. We adopt a mixed quantitative-qualitative methodology (dedicated workshops, interviews with experts, surveys and statistics) with the following specific objectives: i) to facilitate communication and knowledge-sharing across the NBS community by producing a common list of NBS with a nomenclature based on surveyed practitioners preferences; ii) to contribute to the conceptualization of NBS and its mainstreaming by providing a novel and used-friendly classification scheme as well as insights on what defines a solutions as a Nature-based one iii) to overview how the European NBS community evaluates the performance of NBS in terms of UC and ES by providing an integrative assessment based on the qualitative evaluation of each project.

## 2. Methods

The study was divided into 3 major parts: i) development of a common list of NBS (Section 2.1) across H2020 projects through 2 elicitation expert workshops, ii) building consensus on NBS terminology and classification through two worldwide surveys (section 2.2), and iii) development of an integrative assessment framework for UC and ES through statistical analysis of existing scores given in existing H2020 projects (Section 2.3).

## 2.1 Common list of NBS

Our first endeavour was to put together a common list of NBS considering the existing NBS from four European projects. The selected projects were URBANGREENUP (GU), UNALAB (UNL), NATURE4CITIES (N4C) and THINKNATURE (TN), according to the following criteria: (i) funding under H2020 program from at least two different NBS recent calls (2016–2017) (ii) projects should provide an NBS classification scheme including nomenclature and/or description; and (iii) projects should have performed an assessment of NBS in terms of their ability to address urban challenges (UC) and/or provide ecosystem services (ES). We performed a comprehensive review of public documents from these projects (URBANGreenUP, 2018; UNALAB, 2019; NATURE4CITIES, 2020; Somarakis et al., 2019). The initial list of potential NBS was restricted to those with a performance assessment of UC and ES. Then, a face-to-face elicitation workshop (adapted from IDEA protocol, Hemming et al. (2018)) brought together 60 NBS experts during the annual meeting of the COST Action "Circular City" (https: //circular-city.eu/) beginning of March 2020. The workshop was organized in 6 rounds of 4 parallel sessions, each with 15 participants and 1 moderator. The participants randomly changed sessions at each round. The agenda included the following steps:

i. Experts were first asked to reflect on the following questions: what is and what is not an NBS? If it is not an NBS, what is it then? Why is it not an NBS? This conceptual reflection was based on NBS definitions (European Commission, 2015; Cohen-Shacham et al., 2016; Langergraber et al., 2020). Those items not fitting the conceptualization were discarded.

ii. Identifying similar or identical NBS within projects, based on agreed-upon criteria: similar or equal role of nature (natural process occurring); similar or equal technical/design requirements (i.e. elements, costs, materials); similar or equal implementation scale; similar or equal benefits. The NBS that were found to be very similar/ equal were considered as one NBS. Finally, only those NBS considered similar/equal across a minimum of 2 different projects were selected, in order to ensure robustness and to have assessment data from different sources. If needed, a group of NBS experts (within the network of the COST Action –

"Circular city") was consulted to analyse NBS in which the consensus was not achieved. In June 2020 an online elicitation workshop was held to validate procedures for achieving the list of NBS as well as the list itself.

## 2.2 NBS common terminology and classification

Two web-based surveys were launched to collect NBS expert preferences with regard to terminology. In order to ensure a high quality of responses, for both surveys, the respondent could only choose one option and answer only once per survey. The survey was published in OPPLA, the online EU repository of NBS (see https://oppla.eu/nature-based-solutions-terminology-survey; https://oppla.eu/lets-talk-same-nature-based-language-again) and disseminated through other NBS related networks (https://circular-city.eu/, https://snapp.icra.cat/, https://www.edicitnet.com/). In the first survey, a unified description was provided for each NBS based on the information of NBS considered as very similar or equal. The unified description and potential names for the NBS were obtained from the four H2020 projects by consulting public documents (URBANGreenUP, 2018; UNALAB, 2019; NATURE4CITIES, 2020; Somarakis et al., 2019). Respondents were asked to choose between the most suitable names offered for a specific NBS description or to suggest an alternative name. The criterion to select the names was that they received at least 50% of the valid votes (excluding declines and comments). For the cases in which none of the available names for a NBS received at least 50% of the votes, a second survey was launched, which included a thorough description (based on literature and suggestions from the first survey) of the two highest ranking names. For the second survey, the one sample chi-square test based on P-value method with 0.05 level of significance was applied to each NBS to determine if there was a significant difference in the proportion of respondents preferring one of the two names. For those NBS where no clear preference could be established through the statistical analysis of the survey, the final names were defined by the following steps:

i. **Counting the number of publications in Scopus containing the surveyed names** (two most voted names from the surveys). The Scopus search took place in October of 2020. Our criterion was to adopt the name with a minimum of 10 articles. Moreover, names should distinguish NBS from other existing natural ecosystems or other NBS.

ii. **Direct experts' consultation,** only for those NBS in which the name could not be defined in the previous step. In this case, we checked the number of publications in Scopus containing the names suggested by experts against the number of publications containing the surveyed names. The criteria to define the final name was a balance between number of publications and names that properly distinguished a particular NBS from others NBS or natural ecosystems.

A **3-level hierarchical classification** scheme for the NBS was proposed. The **first level** of the classification scheme was built on 3 existing categories proposed by Eggermont et al. (2015) which are: Type 1 represents no or only minimal intervention in ecosystems, with the objectives of maintaining or improving the delivery of a range of ES, both inside and outside of these preserved ecosystems (e.g. protection of mangroves; marine protected areas); Type 2 corresponds to the definition and implementation of management approaches that develop sustainable and multi - functional ecosystems and landscapes, which improves the delivery of selected ES compared to what would be obtained with a more conventional intervention (e.g. planning of agricultural landscapes; enhancing tree species and genetic diversity); Type 3 constitutes the managing of ecosystems in very intrusive ways or even creating new ecosystems (e.g., artificial ecosystems like green walls and green roofs). The fit into these categories was based on literature findings (Almenar et al., 2021 and Sarabi et al., 2019) and experts' interpretations of the Eggermont et al. (2015) categories. The **following two levels** of the new classification were identified by interviewing experts and scanning literature for similarities and differences between the NBS at hand in terms of features, type of vegetation employed, scale of implementation and purpose.

As a general rule, the following criteria were applied to all levels of the classification: 1) an NBS can only fit into one category; 2) all categories have descriptive names (self-standing meaning) and 3) all categories have to be translatable into simple questions as a guidance for using and further developing the classification.

## 2.3 Integrative assessment framework for urban challenges and ecosystem services

Theintegrative assessment of UC and ES was organized in two steps: Calculation and visualization of crossed scores (section 2.3.1) and multivariate analysis (section 2.3.2).

### 2.3.1 Calculation and visualization of crossed scores

Each of the four selected H2020 projects presented a different list of UC and ES. Hence, we first established a baseline for UC, based on the list proposed in the Eklipse framework (Raymond et al., 2017), and for ES based on TEEB (2011) and Millennium ecosystem assessments (Alcamo et al., 2003 and Reid et al., 2005). Next, we related the UC and ES of each project to this baseline. For example, project UNL uses 3 water-related challenges: water scarcity, flood management and water pollution. These challenges were related to the Eklipse challenge "Water management". The same procedure was performed for ES (please see supplementary data: Figure A.1 and Table A.1). This allowed us to establish a common framework for UC and ES assessments. Secondly, we calculated the normalized raw scores for each individual NBS based on the information available in public documents of each project (qualitative approaches: Ecosystem services assessment approach, panel of experts or experts consultation, literature review and etc.) by using binary logic (score 1 when the NBS addressed UC or ES and score 0 when the NBS did not address UC o ES). Next, we calculated the final scores for each of UC and ES. For example, the score for the UNL set of NBS regarding the water management challenge was based on the average of normalized scores for related challenges such as water scarcity, flood management and water pollution. Thirdly, we calculated crossed scores by simply averaging using the normalized raw scores from the NBS considered either as very similar or equal. Finally, a tool was developed using ShinyR package (Chang et al., 2020) on R v. 3.6.3 (R Core Team, 2020) and the ggplot2 package (Wickham, 2016) to visualize the outcome of the UC and ES scores per NBS (find the source-code of the tool in https://github.com/icra/nbs_list).

### 2.3.2 Multivariate analysis

Principal component analysis (PCA) on the UC and ES for all NBS was conducted to explore similarities among NBS and identify which UC and ES were driving those similarities. Data pre-treatment was conducted to avoid biases due to missing values. UC with more than 1 missing value (not assessed for any NBS) were hampered and ES were grouped per type: regulation, provision, cultural and supporting. The package FactoMineR in R (Lê et al., 2008) was used to run the PCA analysis and missMDA package in R (Josse and Husson, 2016) for dealing with missing values. We used the Pielou's evenness index to express which NBS were the most balanced in terms of performance against different facets (UC and type of ES). The evenness index was calculated using the Vegan package in R (Oksanen et al., 2019) as follows:

$$evenness_i = \frac{diversity_i}{\log scores_i} \qquad \text{(Eq. 1)}$$

where *i* is a specific NBS, *diversity* is calculated as a Shannon index and *scores* is the number of variables (crossed scores on UC and ES) in which the *NBS*$_i$ has a score bigger than 0. To calculate the evenness index, the missing values were replaced by zeros.

## 3. Results

First, we present the results of the participatory approach applied to develop the common list of NBS (Section 3.1). Second, we propose a set of names and a novel classification for the NBS included in the list (Section 3.2). Finally, we present the main results of the integrative assessment and multivariate analyses (Section 3.3).

## 3.1 Common list of NBS

The exercise started with more than 250 NBS collected from the four selected H2020 projects. An iterative process was followed to select the ones which i) included an assessment in terms of UC and ES, ii) were considered an NBS, and iii) showed obvious overlapping (similar/equal NBS across different projects) (Figure 1).

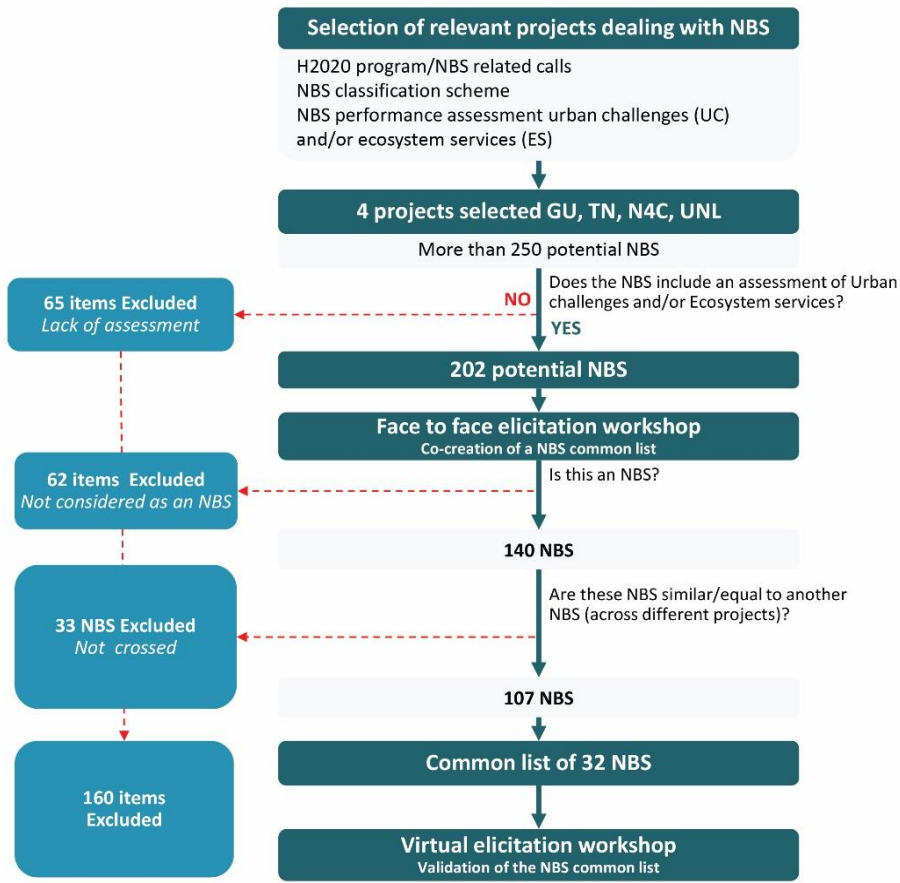

**Figure 1**. Summary of the participatory approach applied to reach the common list of NBS.

As can be seen in Figure 1, a total of 62 potential NBS were excluded for not being considered an NBS. The exclusion of such items was based on the following justifications (see Table A.2): 3 items were found to be a benefit that any NBS could provide, depending mainly on the selection of plant species (e.g. "Cooling trees"); 9 items were found to be inspired by nature, but not employing nature (e.g. "Permeable concrete"); 43 items were found to be supportive planning/management approaches to preserve existing ecosystems and to facilitate NBS implementation, monitoring and its sustainable continuation (e.g. "Limit or prevent some specific uses and practices"); 4 items were found to not be inspired or supported by nature since they were too intensive or did not occur in nature (e.g. "Small-scale urban livestock"); and 3 items were found to be a cluster/category of NBS that could include different NBS with the same purpose (e.g. "Natural wastewater treatment").

Out of the remaining 140 NBS, 33 NBS were excluded because they did not comply with the selection criteria of being similar/equal across at least 2 different projects. Out of the four premises provided to state similarity or equality among NBS, requirements in terms of design and scale implementation proved to be more relevant than benefits or replicating natural processes. Note also that some elements were seen as similar by experts while certain guidelines (University of Arkansas Community Design Center, 2010) consider them as separate NBS (e.g. "Detention ponds" and "Infiltration basins"). Finally, the main outcome of the elicitation workshop is a common list of 32 NBS which represent the 107 NBS from the four H2020 projects reviewed (25 NBS from GU; 30 from UNL; 31 from N4C; and 21 from TN, respectively). Table 1 shows the different names given to the NBS considered similar or equal in each NBS of the H2020.

**Table 1.** Common list of NBS and respective grouped/paired NBS (considered similar or equal across projects).

| NBS | NBS (GROUPED OR PAIRED) | | | |
|---|---|---|---|---|
| | GU | UNL | N4C | TN |
| NBS1 | Floodable park | Infiltration basin; (Dry) Detention Pond | | |
| NBS2 | Grassed swales and water retention ponds | (Wet) Retention Pond | | |
| NBS3 | Rain gardens | Rain gardens | | |
| NBS4 | Grassed swales and water retention ponds | Bioswale | Swale | |
| NBS5 | Electro Wetland | Constructed wetlands | Constructed wetland for wastewater treatment | Use engineered reedbeds/wetlands for tertiary treatment of effluent |
| NBS6 | Green façade with climbing plants | Noise barrier as ground-based greening; Ground-based greening | Climber green wall | Climber green wall |
| NBS7 | Hydroponic green façade; Green noise barriers | Façade-bound greening | Green wall system | Green wall system |
| NBS8 | Vertical mobile garden | Mobile vertical greening / Mobile Green Living Room | | |
| NBS9 | | | Planter green wall | Planter green wall |
| NBS10 | Green shady structures | | Vegetated pergola | |
| NBS11 | Green roof; Green covering shelters | Extensive green roof; Constructed wet roof | Extensive green roof | Intensive green roof/Semi-intensive green roof/Extensive green roof |
| NBS12 | Green roof | Intensive green roof | Intensive green roof | Intensive green roof/Semi-intensive green roof/Extensive green roof |
| NBS13 | | Smart roof | Semi-intensive green roof | Intensive green roof/Semi-intensive green roof/Extensive green roof |
| NBS14 | Natural pollinator`s modules; Compacted pollinator`s modules | | Create and preserve habitats and shelters for biodiversity | Create and preserve habitats and shelters for biodiversity |
| NBS15 | Planting and renewal urban trees; Trees re-naturing parking | Single line trees; Boulevards | Street trees; Single tree | Street trees |
| NBS16 | | Green Corridors | | Green corridors and belts |
| NBS17 | Green resting areas | Residential park | Large urban public park | Large urban park |
| NBS18 | Green resting areas; Parklets | | Pocket garden/park | Pocket garden/park |
| NBS19 | Arboreal areas around urban areas | Group of trees | Wood; Urban forest | Urban forest |
| NBS20 | | | Heritage garden | Heritage park |
| NBS21 | | | Private gardens | Private gardens |
| NBS22 | | | Vegetables gardens | Community garden; Vegetable gardens |

| NBS23 | Urban orchards | | Urban orchard | Urban orchards |
|---|---|---|---|---|
| NBS24 | | | Use of pre-existing vegetation | Use of pre-existing vegetation |
| NBS25 | Community composting | | Composting | |
| NBS26 | Enhanced nutrient managing and releasing soil; Smart soil as substrate | | Soil improvement; Structural soil; Mulching | Soil improvement and conservation measures; Incorporating manure, compost, biosolids, or incorporating crop residues to enhance carbon storage |
| NBS27 | | | Soil & slope revegetation; Strong slope vegetation | Systems for erosion control |
| NBS28 | | Living Fascine; Living revetment; Revetment with cuttings (Spreitlage); Planted embankment mat | Vegetation engineering systems for riverbank erosion control | Systems for erosion control |
| NBS29 | Hard drainage-flood prevention Unearth water courses | Reprofiling the channel cross-section; Channel widening and length extension; Daylighting | Reopened stream | Rivers or streams, including re- meandering, re-opening Blue corridors; Systems for erosion control |
| NBS30 | | Branches; Reprofiling/Extending flood plain area | Floodplain | |
| NBS31 | Hard drainage-flood prevention Unearth water courses | Diverting and deflecting elements | | Systems for erosion control |
| NBS32 | Green parking pavements; Cycle-pedestrian green pavement | Vegetated grid pave | Green parking lot | |

## 3.2 NBS common terminology and classification

The first web-based survey gathered more than 160 participants from 46 countries worldwide and the second survey was answered by 92 participants from 30 countries. For both surveys, most of participants were researchers, engineers, architects, urban planners from Europe (approx. 70%) working in the NBS field for 1-5 years or even more than 5 years. As a result of the first survey, the names of 20 NBS were chosen by following the criteria of more than 50% of total valid votes (Table A.3 – in green). In turn, in the case of 12 NBS, none of the options received most votes (Table A.3 – in blue) and thus, a second web-based survey was carried out. As a result of the second survey, for 7 NBS there was a significant difference (chi-square test, 0.05 level of significance) in the proportion of respondents preferring one of the two options provided (Table A.4 – in green). In contrast, for five NBS no significant difference was noticed: NBS 4, 5, 6, 28 and 30 (Table A.4 - in blue). For NBS 4 and 5, the name selected was the one more commonly used in scientific publications, "Swale" and "constructed wetland", respectively. Moreover, both names can discern such NBSs from others similar NBS or even existing natural ecosystems such as natural constructed wetlands. For NBS 30, even though the name "Floodplain" had a higher number of scientific citations than "Reprofiling/Extending floodplain area", both coming for the survey, the latter name was selected because it distinguishes this NBS from naturally formed floodplains. For NBS 6, the surveyed names "Green façade with climbing plants" and "climber green wal"l were encountered in, respectively 0 and 4 documents in Scopus. Therefore, two options were provided by experts: "Soil-based green façade" and "green façade". No documents were found in Scopus for "Soil-based green façade" and more 236 documents included the name "green façade". Therefore, the later name was adopted because it was encountered in a higher number of articles than the surveyed names and it has been widely applied in literature to differentiate this type of green wall from other typologies (Manso and Castro-Gomes, 2015; Widiastuti et al., 2020; Wang et al., 2020; May Tzuc et al., 2021). For NBS 28, none of the surveyed names have been commonly cited in scientific literature (less than 10 documents were found). Therefore, "Systems for erosion control" was discarded since it does not specify the site in which this intervention will take place, which may cause confusion and difficulties to discern this NBSi from others focused on erosion control. In contrast, "Vegetation engineering systems for riverbank erosion control" and "Riverbank engineering" (suggested by experts) indicate the site of intervention (riverbank). The latter option was selected, since it appeared in a slightly higher number of publications than the other. The results of Scopus search can be seen in supplementary data (Table A.5)

**Table 2.** NBS common list and terminology. The complete version of this table, including description and other possible names suggested through the survey or from existing guidelines, can be seen in Table A.5)

| Approach | Acronym | Name | Approach | Acronym | Name |
|---|---|---|---|---|---|
| 2º survey | NBS1 | *Infiltration basin* | 2º survey | NBS17 | *Large urban park* |
| 1º survey | NBS2 | *(Wet) Retention Pond* | 1º survey | NBS18 | *Pocket garden/park* |
| 1º survey | NBS3 | *Rain garden* | 1º survey | NBS19 | *Urban forest* |
| Mixed[a] | NBS4 | *Swale* | 1º survey | NBS20 | *Heritage garden* |
| Mixed | NBS5 | *Constructed wetland* | 1º survey | NBS21 | *Private gardens* |
| Mixed | NBS6 | *Green facade* | 1º survey | NBS22 | *Community garden* |
| 1º survey | NBS7 | *Green wall system* | 1º survey | NBS23 | *Urban Orchard* |
| 1º survey | NBS8 | *Vertical mobile garden* | 1º survey | NBS24 | *Use of pre-existing vegetation* |
| 1º survey | NBS9 | *Planter green wall* | 1º survey | NBS25 | *Composting* |
| 1º survey | NBS10 | *Vegetated pergola* | 2º survey | NBS26 | *Soil improvement* |
| 1º survey | NBS11 | *Extensive green roof* | 2º survey | NBS27 | *Systems for erosion control* |
| 1º survey | NBS12 | *Intensive green roof* | Mixed | NBS28 | *Riverbank engineering* |
| 1º survey | NBS13 | *Semi-intensive green roof* | 2º survey | NBS29 | *Rivers or streams, including re-meandering, re-opening Blue corridors* |
| 1º survey | NBS14 | *Create and preserve habitats and shelters for biodiversity* | Mixed | NBS30 | *Reprofiling/Extending floodplain area* |
| 1º survey | NBS15 | *Street trees* | 2º survey | NBS31 | *Diverting and deflecting elements* |
| 1º survey | NBS16 | *Green corridors* | 2º survey | NBS32 | *Vegetated grid pave* |

[a] **Mixed: number of scientific publications (Scopus). In the cases in which a Scopus search was not conclusive, literature and experts were consulted (COST Action – circular city). "Circular City").**

The 32 common NBS (Table 2) were classified following a hierarchical structure of 3 levels (Figure 2). The first level distinguishes between **NBS units (NBS$_u$)** (Eggermont et al., 2015 - Type 3) **and NBS interventions (NBS$_i$)** (Eggermont et al., 2015 - Type 2).

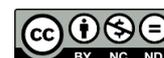

NBS$_u$ are green technologies or green urban spaces, either autonomous or integrated in a larger ensemble - i.e. combined with other NBS, grey infrastructures or conventional technologies - thus forming complex "living" systems. NBSu can be part of existing urban green/blue infrastructure or they can be built from scratch. Finally, these units are capable of replicating processes occurring in nature to enhance the performance of natural capital in cities and thus, provide a wide range of ecosystem services and co-benefits.

NBSi refer to the act of intervening in specific ecosystems or in other NBSu by applying measures or techniques to support natural processes and biodiversity. Even though NBSi can provide diverse co-benefits, they are usually applied to achieve specific purposes (e.g. preserve/maintain natural capital, improve soil quality, prevent/control erosion).

Within the NBS$_u$, on the second level, a differentiation is made between spatial and technological units. **NBS spatial units (NBS$_{su}$)** comprise different elements of urban blue/green infrastructure, mainly related to types of urban green spaces (e.g. parks, gardens, orchards, corridors, forest) or single elements such as street trees. These units are usually implemented on the ground and can provide a plethora of co-benefits. Out of 23 NBS$_u$, 9 were included in this sub-category since they are often referred to as "urban green spaces" (Holt and Borsuk, 2020; Rasli et al., 2019; Nastran, 2020). **NBS technological units (NBS$_{tu}$)** include blue/green technologies that are meant to provide specific features and services (e.g. thermal insulation, shading, water infiltration, water treatment), and thus display a set of specificities in terms of design, implementation and monitoring. In contrast to NBS$_{su}$, NBS$_{tu}$ can be implemented in a wide range of scales, from the ground to vertical empty spaces and rooftops. Moreover, NBS$_{tu}$ can be combined with other NBS$_u$ (spatial or technological) or with other advanced technologies (Maes et al., 2015; Depietri and McPhearson, 2017; Davies and Lafortezza, 2019). Out of 23 NBS$_u$, 14 NBS were assigned as NBS$_{tu}$ as they are often referred to as "urban green *technologies*" or "green *technologies*" or simply "*technologies*" (Stefanakis, 2019; Bonoli and Pulvirenti, 2018; Sun et al., 2018).

On the third level, $NBS_{su}$ and $NBS_{tu}$ are respectively grouped according to forms of vegetation and scale of implementation. $NBS_{su}$ are split in 2 sub-categories: **Spatial Arboreal Units ($NBS_{sau}$),** in which the component arboreal is the main form of vegetation; and **Spatial Mixed Vegetation Unit ($NBS_{smvu}$),** in which a different form of vegetation (beyond trees) can be employed. $NBS_{tu}$ are split in: **Technological Vertical Units ($NBS_{tvu}$),** which are implemented on a variety of vertical surfaces, from façades (of buildings) (Perini, Ottelé and Haas, 2011)(Manso and Castro-Gomes, 2015) to self-standing vertical surfaces anywhere in the city; **Technological Horizontal Units ($NBS_{thu}$)** which are mainly implemented on horizontal surfaces, on the ground or on building rooftops.

On the second level, $NBS_i$ are grouped according to their main purpose or site in which the intervention will take place, resulting in the following sub-categories: **River interventions ($NBS_{ir}$)**, which includes a diverse set of techniques of fluvial/water bioengineering for river management in terms of flow dynamics, flood and erosion control; **Soil interventions ($NBS_{is}$)**, which includes techniques of soil bioengineering to improve and maintain soil quality in terms of physical, chemical and biological features; **Biodiversity interventions ($NBS_{ib}$)** which includes actions and measures for enhancing and preserving the natural capital in cities. The majority of $NBS_i$ were considered river and soil interventions, mainly for being often referred to as soil and water bioengineering techniques (European Federation for Soil and Water Bioengineering, 2015 and Rey et al., 2019). No third level was established for $NBS_i$.

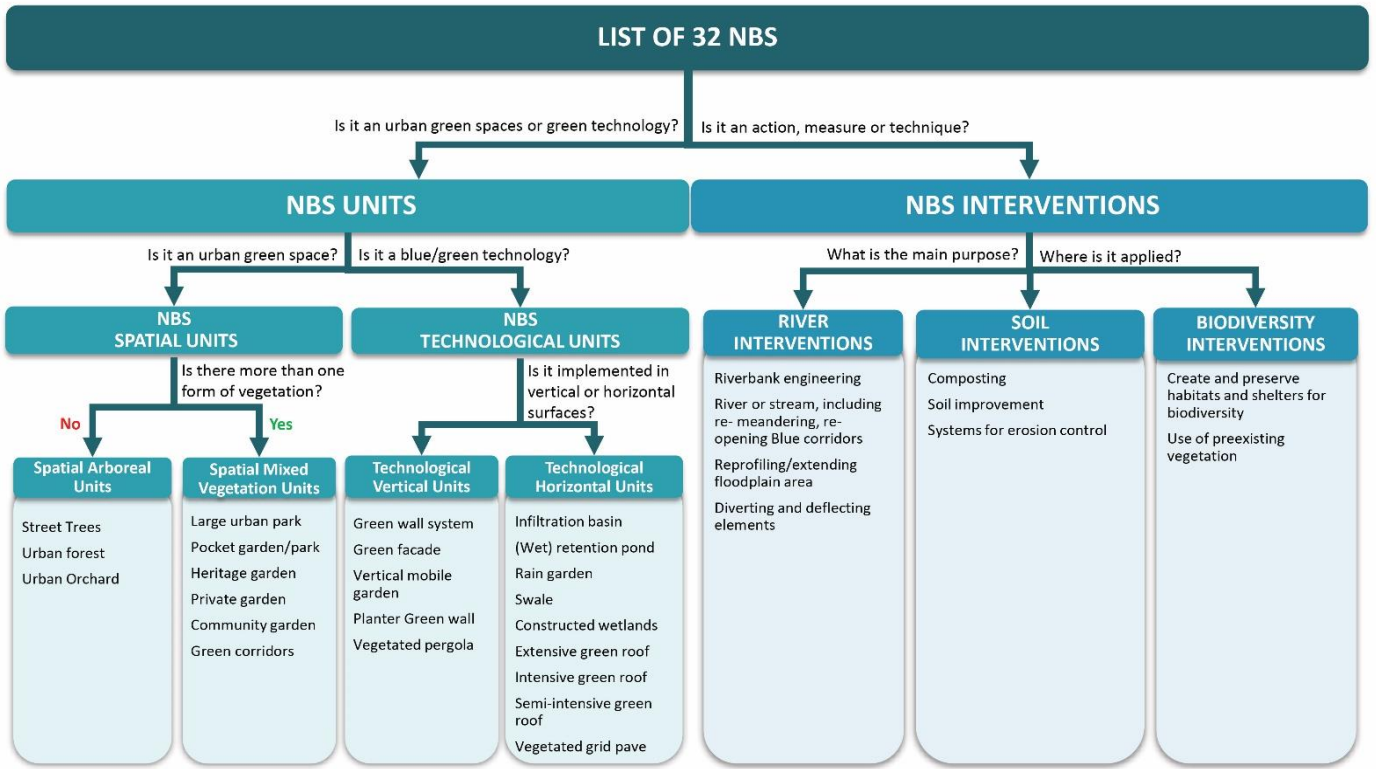

**Figure 2**. A novel hierarchical classification of considered NBS.

### 3.3 Integrative assessment framework for urban challenges and ecosystem services

The integrative assessment framework returned a performance score of NBS between 0 and 1 for 10 UC and 19 ES, with the addition of the four corresponding categories of ES: Regulating, Cultural, Support and Provision. The complete set of results for each NBS can be explored on the tool: https://icra.shinyapps.io/nbs-list. (an example is shown in Figure 3).

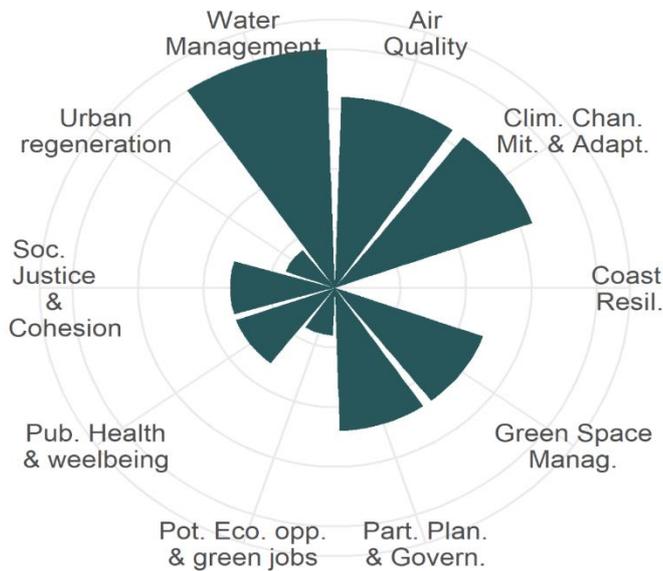

**A. UC addressed by Infiltration basin**

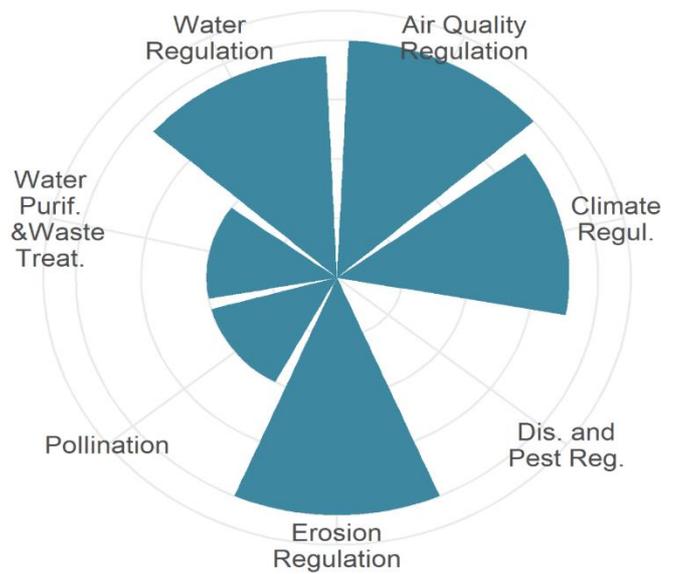

**B. Regulation ES delivered by Large urban park**

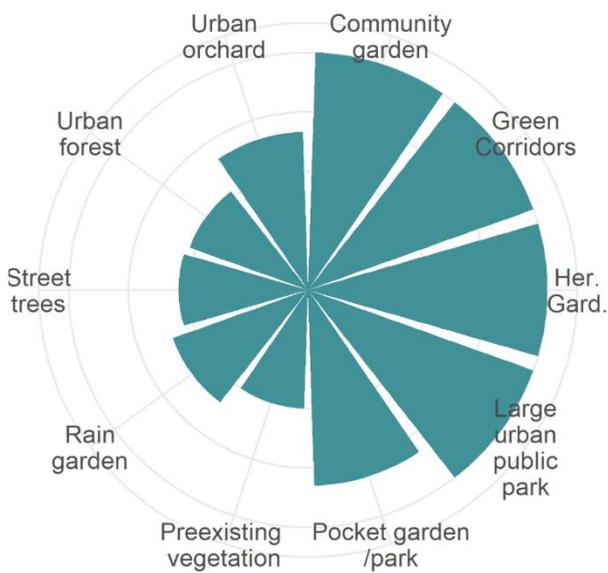

**C. Top 10 NBS addressing social justice**

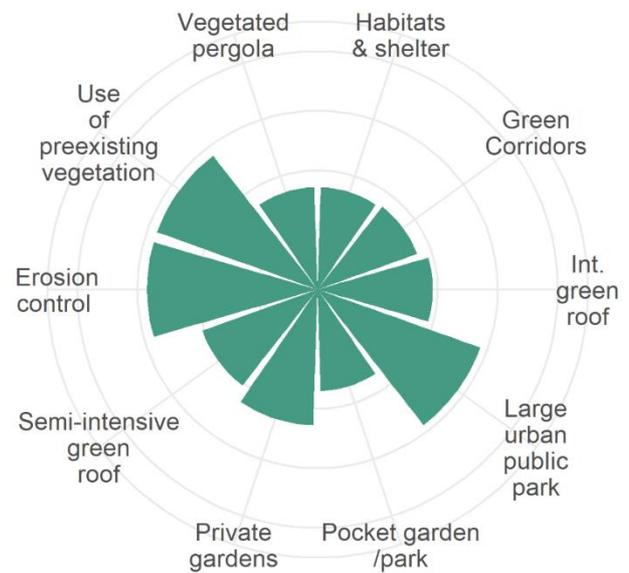

**D. Top 10 NBS addressing regulation ES**

**Figure 3**. Example of plots visualized in the tool.

The distribution of crossed scores through the different assessed challenges and services did not draw a clear pattern (Figure 4). Some UC or ES presented a great diversity while others did not. Green space management is the highest-scoring UC, while public participation is the one with the lowest. Cultural ES is the highest-scoring category of ES, while provisioning's is the lowest-scoring category.

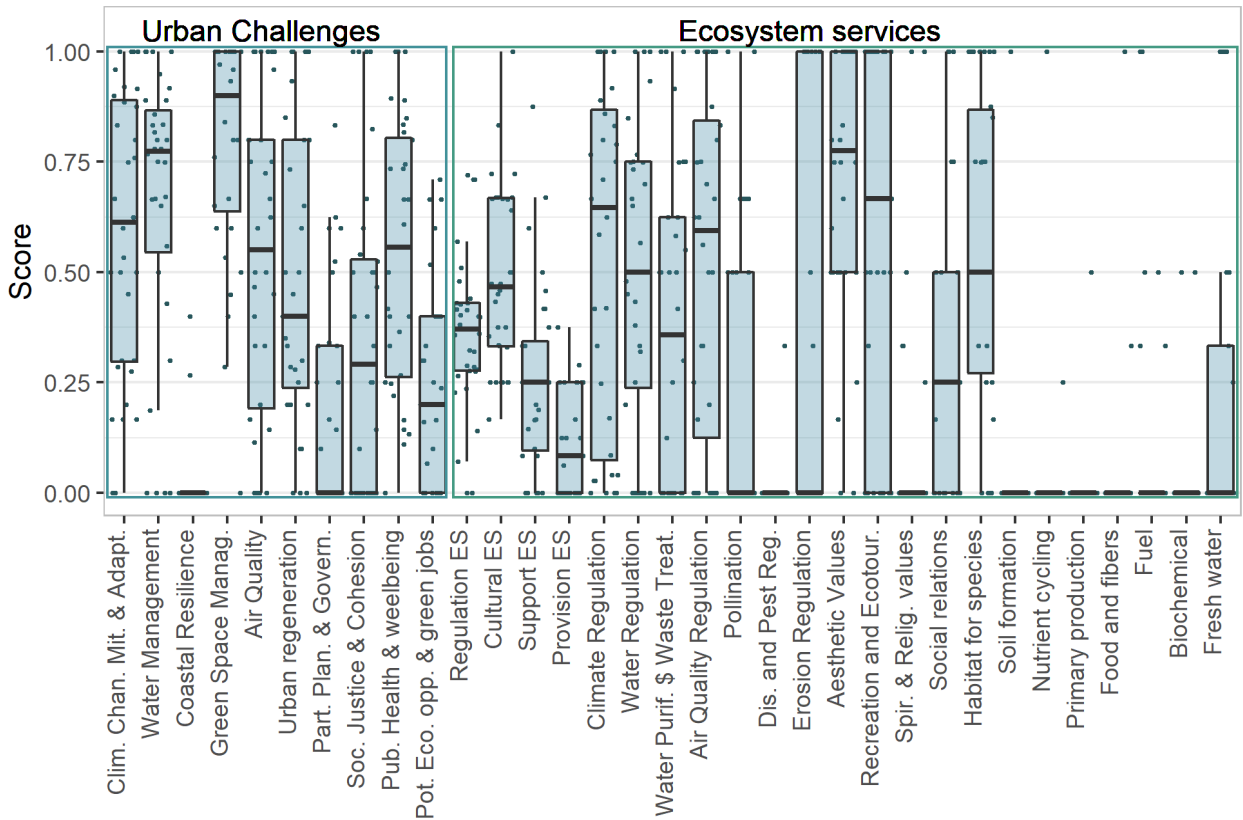

**Figure 4**. NBS crossed scores on urban challenges and on ecosystem services.

The PCA analysis shows that the first two dimensions accumulated 49.4% of model variance (Figure 5). The first dimension is the linear combination of two NBS qualities: social and environmental UC, as well as cultural ES. The second dimension was divided by provisioning and supporting ES and water management in one direction and air quality, climate adaptation and economic opportunities in the other. The PCA outcome shows no obvious clustering of NBS. In addition, the categories of proposed hierarchical classification perform properly in the two dimensions of the PCA, with a few exceptions. Overall, NBS spatial scored well on social issues, climate resilience and air quality; likewise, technological units did not perform too well on water management, green management and cultural ES; finally, most of interventional NBS performed well in provision and supporting ES.

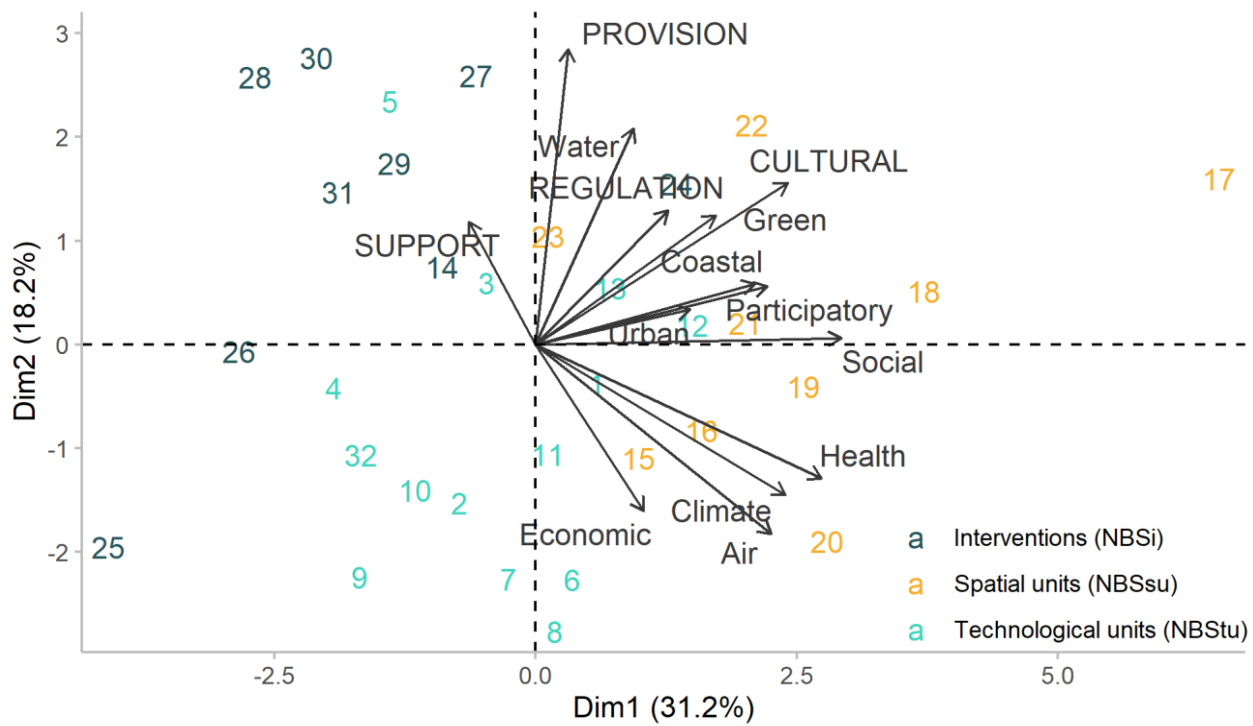

**Figure 5**. Principal Component Analysis of the considered NBS, ordinated as a function of their scores on urban challenges (lower-case) and ecosystem services (upper-case), and colored as a function of the hierarchical classification.

The evenness index for all NBS is shown in Figure 6. All NBS scored between 0.85 and 1, thus, offering good overall performance addressing UC and providing ES. In general, $NBS_u$ presented higher scores in the evenness index than NBSi, showing that spatial and technological units tend to be more generalist, whereas NBSi suit better for addressing specific UC and providing specific ES.

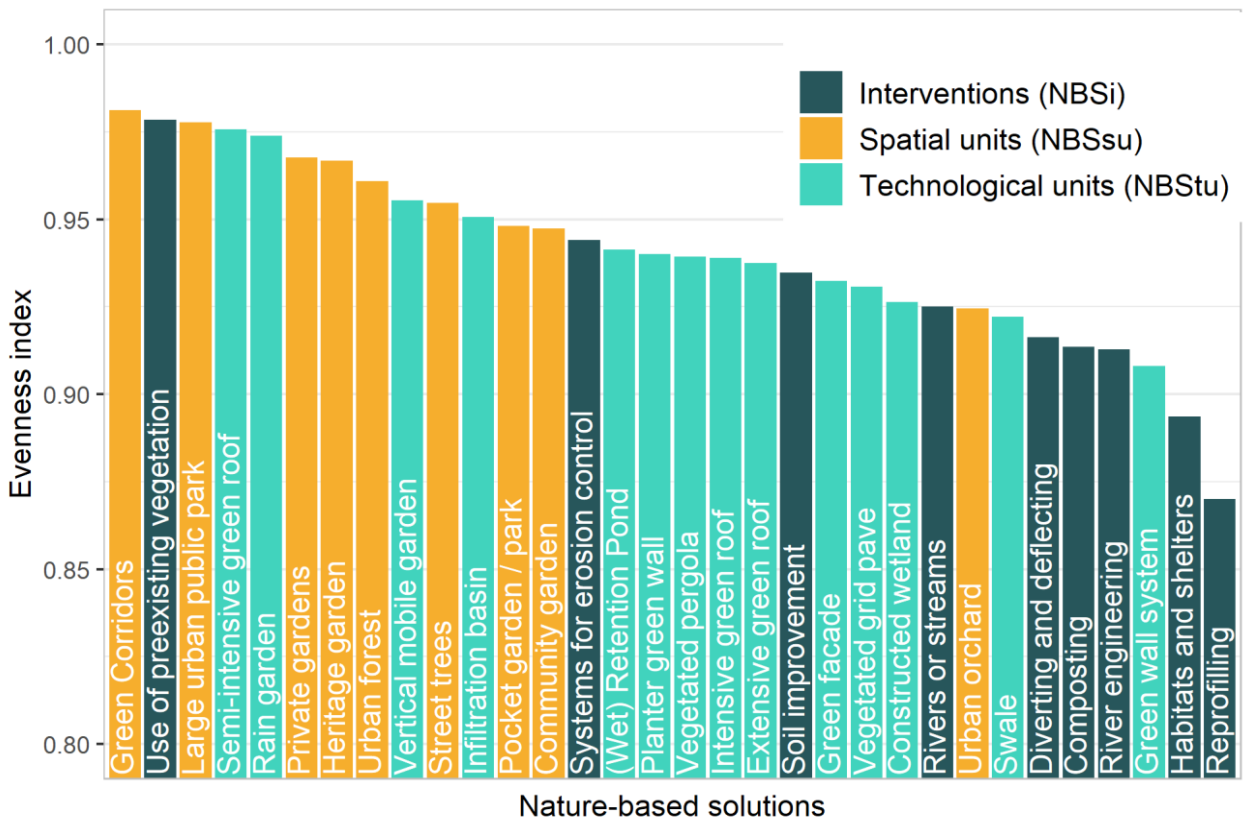

**Figure 6**. Evenness index for the considered NBS, coloured as a function of the second level of the hierarchical classification.

## 4 Discussion

This section puts forward a critical discussion about criteria raised during workshops to state if a solution is based on nature or not (section 4.1). Next, we discuss the importance of having a common understanding of NBS terminology and classification in order to foment a path towards standardization (Section 4.2). Finally, we present a discussion about the integrative performance assessment (section 4.3)

### 4.1 To be or not to be an NBS: What does it take?

The participatory approach employed in this research has revealed a set of insights on why a solution can be considered (as) Nature-Based one. The term 'nature' is key in this context: according to Dorst et al. (2019)

one of the principles of NBS is that "*nature, as the concept's central foundation, may take many forms*". Indeed, the understanding of what constitutes 'nature' has caused disagreements and intense scholarly debates, especially in the field of political ecology (Robbins, 2012; Kotsila et al., 2020). Consequently, this makes it challenging to define what can be understood as 'nature' within the scope of NBS (Carsten et al., 2017). In this sense, our results suggest that 'nature' is often understood as - what we would call – the 'green factor', defined here as the presence of vegetation. This is indirectly sustained by the IUCN definition, in which NBS are expected to provide biodiversity benefits (Cohen-Shacham et al., 2017). The '**green factor´** plays an important role when deciding if a certain solution is a Nature-Based one or not. For example, participants (workshops and survey) often associated "being inspired in nature" with the replication of natural processes, as sustained by European Commision (2015) definition. Yet, this was a necessary but not sufficient condition to consider a particular item an NBS: the item in question also had to **employ nature**. For instance, "porous asphalt", replicates natural process of water infiltration, yet it does not necessitate the presence of the green factor.

Another example emphasizing the relevance of the 'green factor´ is that all NBS units included some form of vegetation, while approximately 70% of them mentioned green-related terms in their names (e.g. "green", "vegetated", "garden", "forest"). This trend could be interpreted almost as an unquestionable, unconscious frame: in order to count as an NBS unit, a solution must be green. In the specialized literature, NBS are often related to expressions such as "greening of cities" (Tozer et al., 2020), "urban green space" (Panno et al., 2017), "urban greening" (Dorst et al., 2019 and Escobedo et al., 2018), "greening strategy" (Fastenrath et al., 2020) and "green placemaking" (Gulsrud et al., 2018). This frame can be explained by the need to distinguish the natural foundation of NBS in comparison to other conventional grey infrastructure. Moreover, there are current efforts to make cities 'greener'. According to Kotsila et al. (2020) "*in Europe,* *specifically, NBS are seen not only as an alternative means to address social needs and enhance natural environments but also as a way of **boosting green innovation** and resilience in cities*".

Apart from the replication and employment of 'nature', other criteria emerging from the participative process were the following: 1) an NBS should be **non-intensive** (in terms of resources) (European Commision, 2015; Faivre et al., 2017) and 2) and NBS should **occur in nature** (Frantzeskaki, 2019)**.** Based on this, items like "Smart soil production in climate-smart urban farming precinct" and "Small-scale urban livestock" have not been considered NBS: the former is not found in nature and the latter can involve intensive use of resources.

It is important to note that by requiring the fulfillment of the factor 'green' for an element to be considered as NBS, we eliminate a whole range of elements based on (bio)filtration through natural porous material. A good example of this is an infiltration trench commonly used for sustainable urban drainage. The role of natural, non-intensive processes (filtration and/or biofiltration) is evident, yet the requirement for 'greening' is not fulfilled.

Moving forward, our results also reveal a resistance to accepting planning/management approaches as NBS. Also, while there was a clear consensus among participants about not including these items, they did not convey a clear justification for this. Yet, the literature validates the criteria of the participants: planning/management approaches belong to the types 1 and 2 of Eggermont et al. (2015), which according to Sarabi et al. (2019) were rarely recognized as NBS. Moreover, according to Almenar et al. (2021), there is currently a demand to bring palpable natural structures back to the cities, because they are perceived as more effective than the solutions focused on managing and restoring. Thus, it should not come as a surprise that consulted experts are more inclined to associate NBS with tangible structures rather than more abstract concepts such as management and planning approaches. Nevertheless, we suggest to consider such planning and management aproaches as supportive actions or elements, mainly due to their great relevance regarding the preservation/maintenance of natural capital in cities and implementation and monitoring of NBS.

## 4.2 The path to NBS standardization

Recently the IUCN released the "Global Standard for Nature-based Solutions", focused on design and upscaling (IUCN, 2020). This is much-needed progress in terms of facilitating NBS implementation and mainstreaming. Nevertheless, it is equally important to couple these standards with a common set of NBS based on a formalized terminology.  In this sense, our list represents an advancement in comparison to the lists provided by the 4 analyzed H2020 projects for two reasons. Firstly, it identifies similar NBS among 4 different European projects, which constitute a 'core' set of NBS across all projects. Secondly, it relies on the opinion of worldwide experts within and beyond the EU (e.g. H2020 projects, COST Action "Circular City" or OPPLA community) regarding what an NBS is and how to name it. This presented NBS list is, therefore, an important step towards NBS standardization and it has the potential to evolve in time, with future advances of NBS concept.

Another key advancement of this research is the proposal of a novel NBS classification scheme. Each of the projects analyzed here presented different classifications. Some of the categories proposed in these projects were too broad and thus some NBS could fit in more than one category. For example, the UNL category "greening interventions" – which, strangely, only includes interventions containing trees as the main element - clearly overlaps with other categories such as "vertical greening" and "public green space". In turn, GU properly differentiates the "arboreal interventions" from other types of greening interventions, even though it includes categories such as "carbon capture" with only one item:  planting trees for carbon sequestration. Moreover, the categories proposed by TN provide no clear guidance, apart from the name of the category itself and the conceptual definition it entails. N4C represents a step forward with respect to the previous classifications since it develops a comprehensive hierarchical classification with more than 20 categories. However, some inconsistencies were identified. For example, some NBS such as "Green walls" and "Vegetated pergola", included in the building category, can be also applied in other sites not strictly related to a building surface (e.g self-standing structures or self-standing walls). In addition, some NBS included in the category "on building & structures", such as green roofs and green walls, could be also included in the category "Water" since they can be designed for water retention and treatment. While these classifications are valuable in terms of putting much needed order in a burgeoning field, we believe there is room for a more accessible, user-friendly and all together simpler way to classify NBS. Therefore, to bridge these gaps, we propose a more compact hierarchical classification, with only 11 categories which are conceptually fine-tuned. Additionally, simple questions support the classification into these categories, which are meant to guide practitioners to select the most suitable NBS according to their needs or to further develop the classification by either, including other NBS in our classification scheme or creating new categories.

Regarding the terminology proposed in this article, this research offers important insights about how people perceived NBS trough the way they name it. In this sense, the names of $NBS_{su}$ contained terms referring to its location, scale/size, ownership and type of structure (e.g. **street** trees, **large or pocket** parks, **private or community** gardens). Almost all names of $NBS_{tu}$ also included terms related to the type of structure to be greened (e.g. green **wall** system, green **roof**, vegetated **pergola**). Some $NBS_{tu}$ contained terms to distinguishing them from existing natural ecosystems (e.g. c**onstructed** wetlands), to highlight the process occurring (e.g. **infiltration** basin, (wet) **retention** Pond) or to characterize design requirements (e.g. **intensive, extensive** and, **semi-intensive** green roofs). In contrast, the names of $NBS_i$ , in general, included verbs referring to the actions to be applied, fact which helps to discern the NBSi from existing ecosystems and other NBSi applied in the same context or with similar purposes (e.g. re-meandering, diverting, reprofiling, composting, create, preserve, use).

Furthermore, even if the majority of NBS were named as a result of the surveys, it is undeniable that there is still a lack of agreement within the NBS community. While the survey participants proposed more than 250 new names, others made mention of the existence of well-established guidelines that included accepted terminologies for several of the NBS listed (http://www.efib.org/; https://boku.ac.at/baunat/iblb; https://www.cirf.org/en/home-9/, Woods-Ballard et al., 2015; FFL, 2018; University of Arkansas community design center, 2010). This contradiction can be explained by the fact that terminologies are formulated under social, ethnic and cognitive criteria in which communication among experts and specialists can produce different terms for the same concept and more than one concept for the same term (Faber and Lopez-Rodríguez, 2012). Therefore, as concepts and terms tend to evolve over time, we consider that the application of cognitive models coupled with the reviewing of existing standards in the field of NBS terminology could be helpful to validate the terminologies proposed in this article.

## 4.3 Assessment for urban challenges and ecosystem services

Integrating qualitative assessments from different projects puts forward an overview of how the European NBS community (experts, urban planners and other practitioners) evaluates the performance of NBS in addressing UC and ES. Our results indicate that the impact of NBS on environmental challenges such as "Climate mitigation and adaptation" ($M_e$ = 0.61), "Water management" ($M_e$ = 0.77) and "green space management" ($M_e$ = 0.90) might be perceived as more relevant than social-economic challenges such as "Participatory planning" ($M_e$ = 0.0), "Social justice and cohesion" ($M_e$ = 0.29) and "Economic opportunities" (0.20). The same can be seen in terms of "Regulation" ($M_e$ = 0.37) and "Cultural services" ($M_e$ = 0.47) which received higher average scores than "Supporting services" ($M_e$ = 0.25) and "Provisioning services" ($M_e$ =

0.08). Except for "Habitat for species" and "Fresh water", all remaining supporting and provisioning services received a score of 0 for more than 90% of NBS. This might suggest that even though NBS are multifunctional, there is a greater agreement regarding the role of NBS in addressing environmental challenges, regulating and cultural services than in what concerns social-economic oriented challenges and supporting and provision services. The disparity is likely not indicative of the NBS potential to address UC or provide ES, but rather reflective of uneven efforts in the scientific community in evaluating NBS impacts on said challenges and services. In this sense, the review performed by Almenar et al. (2021) reveals that the great majority of scientific articles are dealing with NBS related to water management and climate change, while only very few papers address social challenges such as public participation and governance.

Regarding the multivariate analysis, the PCA and the evenness index showed that $NBS_{su}$ are providing more co-benefits than $NBS_{tu}$ and $NBS_i$. This may be because $NBS_{su}$ have been more explored by researchers than other typologies and, consequently, the co-benefits are better documented (Almenar et al., 2021). Nonetheless, there are two important insights of the multivariate analysis: (1) there is no overlap in the final list of NBS in terms of UC and ES, that is, each NBS in the list is useful for different situations, and (2) the classification of NBS, despite being designed in terms of visual and functional aspects, works well in terms of addressed UC and delivered ES. This shows the robustness of the proposed classification scheme.

All explored projects edited NBS catalogues in static format (such as pdf documents). In addition, others like N4C or GU presented more dynamic tools to support decision making such as the "Nature Based Solutions explorer" of N4C (https://nbs-explorer.nature4cities-platfor m.eu/) and the NBS selection tool of GU (https://www.urbangreenup.eu/resources/nbs-selection-tool/nbs-selection-tool. kl). However, as far as we know, there is no tool quantitatively evaluating the performance of NBS both in terms of addressing UC and delivering ES. The tool presented in this article (https://icra.shinyapps.io/nbs-list) provides the scores of each NBS in terms of UC, ES and the subsequent categories. More precisely, it provides the scores visualization of a specific NBS (Fig. 3 A and B), along with its description, as well as the visualization of different NBS' scores regarding an individual UC or ES (Fig. 3 C and D). This allows experts and practitioners to explore the co-benefits of any NBS or identify the NBS that best addresses a specific issue.

## 5  Conclusions

There is still a long way to reduce the existing gap between NBS technological development (scientific community), practical and cost-effective applications (public-private sectors) and existing values of the civil society. It is not an easy gap to address, especially when there is no common agreement on the NBS conceptualization.  In this sense, our research represents the perspectives of a wide and diverse NBS community and thus contributes to set a path towards common understanding of the NBS concept.

This article proposes a replicable methodology which delivers a set of 32 NBS fully evaluated in terms of UC and ES and a novel classification scheme, robust enough for systematic knowledge representation and open for further expansion. Such efforts can enable further integration of databases beyond the scope of the four H2020 projects analysed here. Moreover, the proposed classification scheme represents a step forward in the conceptualization of NBS: they are no longer seen as sole elements but as part of large ensembles, forming complex "living" systems (e.g. NBS combined with other NBS or grey infrastructure or conventional technologies). In this sense, we believe that NBS can gain from a more holistic perspective, in which their interactions with other solutions (Nature-Based or not) can help diversify the provision of ES, close the resources loops, and compensate possible disservices.

The results indicate that the 'green factor' and non-intensive (in terms of resource use) solutions occurring in nature are key aspects for practitioners to identify a particular solution as an NBS. Such insights can facilitate the improvement of existing NBS definitions towards clearer and, at the same time, more comprehensive ones. However, more research coupling cognitive analysis with other qualitative methods could shed light on why people tend to privilege certain terms over others when naming NBS. Such mixed quantitative-qualitative approaches can also provide a better understanding of 'nature' as an 'empty signifier' - e.g. a term that can display a set of different, sometimes even opposing meanings (Brown, 2015) in the scope of NBS, thus engaging with current debates in political ecology on the matter.

The common NBS list, terminology and classification scheme proposed in this article are not definitive but intended to evolve in time, along with relevant advancements in the NBS field. Further research is needed in order to discern between NBS interventions – as defined in our classification - and other green planning/management approaches. It is important to better understand, for instance, why practitioners excluded many of the latter. Moreover, it is important to consider, if these items had been considered, what would this have entailed for the current overall conceptualization and classification. Indeed, some of the excluded NBS can be further reviewed (especially those excluded for not having a performance assessment of UC and ES or those that were not found in more than one project).

The overall results of the integrative assessment suggest a need to enhance scientific efforts in evaluating NBS performance in terms of socio-economic challenges, supporting and provisioning services, especially in the case of technological and interventional NBS.

All in all, this article defends that NBS community (research, public and private sector) needs to speak the same "Nature-Based" language in order to further facilitate the knowledge transfer, replication and the engagement of civil society and thus promote a real change in the way citizens perceive the relevance of NBS for making their cities more resilient.

**Acknowledgements**

Authors acknowledge the support from the European Union Horizon 2020 research project EdiCitNet (GA776665), the COST Action Circular City (CA 17133 ), CLEaN-TOUR project (CTM2017-85385-C2-1-R), the Economy and Knowledge Department of the Catalan Government through Consolidated Research Groups

ICRA-ENV (2017 SGR 1124) and ICRAtech (2017 SGR 1318). ICRA researchers thank funding from CERCA program (Generalitat de Catalunya). Lluís Corominas acknowledge the Ministry of Economy and Competitiveness for the Ramon and Cajal grant and the corresponding I3 consolidation (RYC-2013-14595). Nataša Atanasova acknowledges the project "Closing material flows by wastewater treatment with green technologies" (ID J2-8162), financially supported by the Slovenian Research Agency. All authors acknowledge the great contribution of Katie Kearney concerning language refinement. The authors would like to give a special thanks to James Atkinson from OPPLA for supporting the diffusion of our surveys. Finally, the authors also acknowledge all participants of the online surveys for their great contributions concerning the definition of the NBS terminologies.

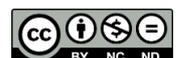

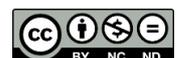

Supplementary data

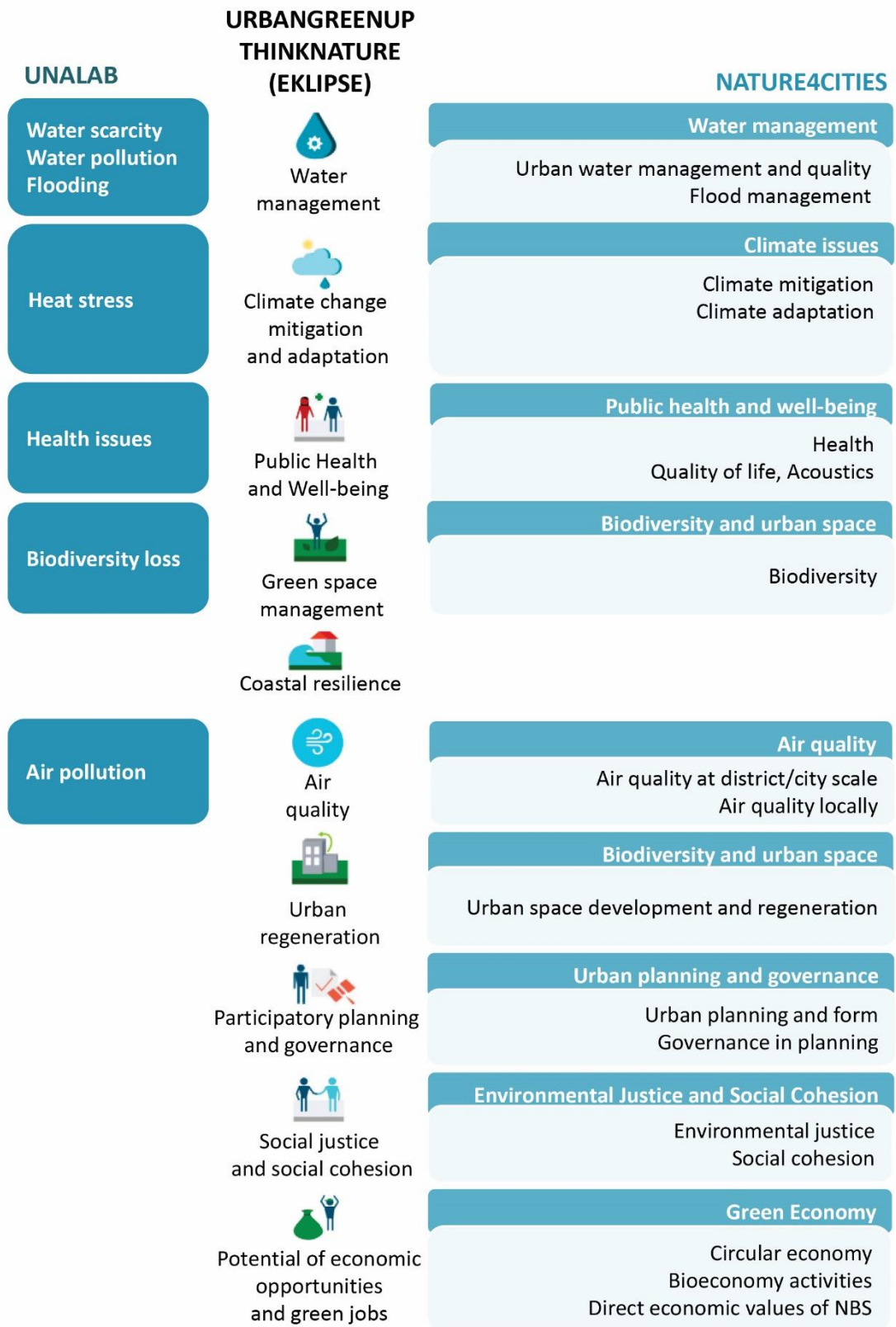

**UNALAB**

**URBANGREENUP THINKNATURE (EKLIPSE)**

**NATURE4CITIES**

Water scarcity
Water pollution
Flooding

Water management

Water management

Urban water management and quality
Flood management

Heat stress

Climate change mitigation and adaptation

Climate issues

Climate mitigation
Climate adaptation

Health issues

Public Health and Well-being

Public health and well-being

Health
Quality of life, Acoustics

Biodiversity loss

Green space management

Biodiversity and urban space

Biodiversity

Coastal resilience

Air pollution

Air quality

Air quality

Air quality at district/city scale
Air quality locally

Urban regeneration

Biodiversity and urban space

Urban space development and regeneration

Participatory planning and governance

Urban planning and governance

Urban planning and form
Governance in planning

Social justice and social cohesion

Environmental Justice and Social Cohesion

Environmental justice
Social cohesion

Potential of economic opportunities and green jobs

Green Economy

Circular economy
Bioeconomy activities
Direct economic values of NBS

Figure A.1 Common baseline of Urban challenges addressed by NBS. (adapted from Raymond et al., 2017; NATURE4CITIES, 2020; UNALAB, 2019; URBANGREENUP, 2018; Somarakis et al., 2019)

Table A.1 Common baseline of ecosystem services provided by NBS.

| | [1]ECOSYSTEM SERVICES | GU | TN | UNL |
|---|---|---|---|---|
| **PROVISIONING services** | *Food and fibers* | Food and Fibers | Food, crops. Wild food and spices | X |
| | *Fuel* | Fuel | Energy | X |
| | *Biochemicals, natural medicines, and pharmaceuticals* | Biochemicals, natural medicines, and pharmaceuticals | Pharmaceuticals, biochemicals, and industry products | X |
| | *Fresh water* | Fresh water | Water | X |
| **REGULATION SERVICES** | *Climate regulation* | Climate regulation | Carbon sequestration and climate regulation. (CS&R) | *Colling service:* Transpiration; Evaporation; Shading; Insulation building |
| | *Water regulation* | Water regulation | Flood protection | *Water regulation service:* Conveyance; Infiltration; Retention; Storage; Reuse |
| | *Water purification and waste treatment* | Water purification and waste treatment | Water purification | *Water purification service:* Filtering; Biofiltration |
| | *Air quality regulation* | Air quality maintenance | Air quality regulation | *Air purification service:* Deposition; Filtration |
| | *Erosion regulation* | Erosion control | Erosion Prevention | X |
| | *Pollination* | Pollination | Crop pollination | x |
| | *Disease and pest regulation* | Regulation of human diseases | Pest and disease control | |
| **CULTURAL SERVICES** | *Aesthetic value* | Aesthetic value | Intellectual and aesthetic appreciation | *Amenity value service:* Beauty appearance |
| | *Recreation and ecotourism* | Recreation and ecotourism | Recreation | x |
| | *Social relations* | Social relations | x | *Amenity value service:* Social interaction |
| | *Spiritual and religious values* | Spiritual and religious value | Spiritual and symbolic appreciation | x |
| **SUPPORTING services** | *Habitats for species* | x | Maintaining populations and habitats | *Biodiversity:* Habitat provision; Connectivity |
| | *Soil formation* | Soil formation | Soil formation and composting | x |
| | *Nutrient cycling* | Nutrient cycling | Nutrient dispersal and cycling | x |
| | *Primary production* | Primary production | Primary production | x |

[1]Adapted from (Mader et al., 2011) and Millennium ecosystem assessment (Alcamo et al., 2003)(Reid et al., 2005)

Table A.2 – Excluded Items/NBS

| MOTIVE | PROJECTS | | | |
|---|---|---|---|---|
| | URBANGREENUP (2018) | UNALAB (2019) | NATURE4CITIES (2020) | THINKNATURE |
| *Lack of performance assessment in terms of UC and ES* | | | Hedge and planted fence; Flower field; Meadow; Community garden; Quarry restoration; Plant with bio-filter features, Excavation of new waterbody (pond, lake); Infrastructure removed on river (ex. Dam); Remeander river; Re-profiling river bank; Use of grazing animals; Revegetation of aquatic planting; Gravity fountain (captation of a spring); Constructed wetland for phytoremediation; Use of terrace (based on cultivation terraces principles), Roof pond; Insect hotel (for wild bee); Beehive for honeybees | Mulching; Integrate biochar into agricultural soils; Hedge and planted fence; Flower strips; Cover crops; Wind breaks; Encourage development of early successional sand dune habitats (dry dunes and wet slacks) where carbon sequestration rates are high; Enhance or facilitate habitat expansion, including the facilitated range expansion of mangroves, as warming conditions and changes in storm occurrence permit; Historical urban green network structure; Urban natural protected areas; Introduced vs. local plants; Vegetation diversification; Mapping green features; Integrated pest management; Integrated weed management; Integrated and ecological management - time and frequency aspects; Green cemetery; Hedge and planted fence; Flower field; Roof Pond; Urban vineyards; Meadow; Urban farm; Introduced vs. local plants; Vegetation diversification; Plant and bio-filter features; Moss green roofs; Direct human intervention; Planning tools for climate change adaptation/mitigation and ecosystem services; Quarry restoration; Phytoremediation; Soil and slope revegetation; Strong slope revegetation; Replace hard engineered river stabilization with softer alternatives (e.g. willow-based); Plant trees/ hedges/perennial grass strips to intercept surface run-off; Restore wetlands in areas of groundwater recharge; Reconnect rivers with floodplains to enhance natural water storage; Re-vegetation of riverbanks; Restore grassland/low input arable in drinking water catchments; Target ponds/wetland creation to trap sediment/pollution runoff in farmed |

| | | | |
|---|---|---|---|
| | landscape; Floodplain restoration and management; Reshape river and riverbanks in urban areas; Create new intertidal habitat through afforestation, or planting of saltmarsh or seagrass at appropriate elevations in the tidal frame; Restore micro-topography, creek networks, sediment inputs, and nutrient exchange in abandoned aquaculture ponds; Ecological restoration of degraded coastal and marine ecosystems; Coastal sand engine; Dune replenishment | | |
| ***Not an NBS, but a benefit or a function that depends on selection of plants species*** | Shade trees; Cooling trees; Urban carbon sink | | |
| ***Inspired by nature but not employing nature*** | Hard drainage pavements; Cool pavement | Permeable pavement; Permeable concrete; Porous asphalt; Permeable stone carpet; Underground water storage | Unsealed parking lot; De-sealed area |
| ***Not an NBS. It is Too intensive and/or does not*** | Smart soil production in climate-smart | Lawn | |
| *happen in nature itself.* | urban farming precinct; Climate-smart greenhouses; Small-scale urban livestock | | |
| *Not an NBS, but a typology of NBS* | SUDS; Natural wastewater treatment | Sustainable Urban Drainage Systems | |
| *Not an NBS, but a planning approach or tool to support implementation of NBS or to preserve/monitor natural ecosystems and existing NBS* | | Sustainable use of fertilizers; Integrated pest management; Integrated **weed** management; Integrated and Ecological Management: spatial aspect; Integrated and Ecological Management: time and frequency aspects; Limit or Prevent Access to an Area; Limit or prevent some specific uses and practices; Ensure continuity with ecological network; Take into account the distributions of **public green spaces** through the city; Planning | Limit or prevent specific uses and practices; Protect forests from clearing and degradation from logging, fire, and unsustainable levels of non-timber resource extraction; Maintain and enhance natural wetlands; Protect remaining intertidal muds, saltmarshes and mangrove communities, seagrass beds, and vegetated dunes from further degradation, fragmentation, and loss; Natural Protected Area network structure; Mangrove forests protected area; MPA network structure; Agro-ecological practices; Use grazing management and animal impact as farm and ecosystem development tools; Change crop rotations; Agro-ecological network structure; Enrichment planting in degraded and regenerating forests; Increase soil water holding capacity and infiltration rates; Deep-rooted plants and minimum or conservation tillage; Integrated coastal zone management; Ensure continuity with ecological network; Planning tools to control urban expansion; Planning tools for biodiversity, green infrastructure, and ecosystem services; Tools to engage citizens; Assessment of NBS benefits; Ecosystem services valuation methods; Bio-indicators; Integrated and ecological management - spatial aspects; Choices of plants; Urban network structures; Use of |

| | | | | |
|---|---|---|---|---|
| | | | tools to control urban expansion; Bio-indicator | fauna; Account for distribution of public green spaces through the city; Mapping of urban green connectivity and biodiversity; Develop urban blue infrastructure; Integrated water management; Re-establish and restore previous intertidal habitat by de-poldering or coastal realignment |
| ***NBS are Not similar/equal across different projects*** | Cycle and pedestrian green route; Green fences; Floating gardens; Green filter area; Channel re-naturing; Urban catchment forestry; Urban garden bio-filter; Green filter area; Pollinator verges and spaces; Pollinators walls/vertical; Pollinator roofs | Free standing living wall; Noise barrier as free-standing living wall; Moss wall ‚City tree' ; Living Plant Constructions (Baubotanik); Biofilter (water purification); Biofilter (air purification); Mounds; | Botanical garden; Green cemetery; Public urban green spaces (Squares. Neighbourhood green space amenity green space); Urban green space with specific uses (school. playground. camp ground. sport field); Green tram track; green strip; green waterfront; Urban vineyard; Urban farm; Phytoremediation; Introduced plants; Vegetation diversification; | Constructed wetlands and built structures for water management; Agroforestry; Forest patches |

Table A.3 Result of the First web-based survey. In green selected name with more than 50% of valid votes. In blue the two most voted names used in the second web-based survey (obs.: The table shows only names with a minimum number of votes following the equation: nº total valid votes/number of possible name, including default options and name suggested by participants; The option "others" includes all names with less than the minimum number of votes)

| | NAME | % | | NAME | % | | NAME | % |
|---|---|---|---|---|---|---|---|---|
| **NBS1** | **Infiltration basin** | **41.1** | **NBS2** | **(Wet) Retention Pond** | **63.7** | **NBS3** | **Rain garden** | **88.6** |
| | **(Dry) Detention Pond** | **34.2** | | Grassed swales and water retention ponds | 27.4 | | Others | 11.4 |
| | Floodable park | 17.1 | | Others | 8.9 | | | |
| | Others | 7.5 | | | | | | |
| **NBS4** | **Bioswale** | **42.2** | **NBS5** | **Constructed wetlands** | **48.6** | **NBS6** | **Green facade with climbing plants** | **41.8** |
| | **Swale** | **38.3** | | **Constructed wetland for water treatment** | **35.1** | | **Climber green wall** | **27.4** |
| | Grassed swales and water retention ponds | 15.6 | | Others | 16.2 | | Green wall | 8.9 |
| | Others | 3.9 | | | | | Noise barrier as ground-based greening | 6.8 |
| | | | | | | | Others | 15.1 |
| **NBS7** | **Green wall system** | **68.6** | **NBS8** | **Vertical mobile garden** | **65** | **NBS9** | **Planter green wall** | **72.2** |
| | Hydroponic green facade | 7.9 | | Mobile vertical greening / Mobile Green Living Room | 21.7 | | Others | 27.8 |
| | Façade-bound greening | 7.9 | | Others | 13.3 | | | |
| | Others | 15.7 | | | | | | |
| **NBS 10** | **Vegetated pergola** | **56.3** | **NBS11** | **Extensive green roof** | **53.8** | **NBS12** | **Intensive green roof** | **60.4** |
| | Green shady structures | 30.4 | | Green roof | 33.3 | | Green roof | 25.5 |
| | Others | 13.4 | | Others | 12.8 | | Others | 14.1 |
| **NBS13** | **Semi-intensive green roof** | **63.5** | **NBS14** | **Create and preserve habitats and shelters for biodiversity** | **53.8** | **NBS15** | **Street trees** | **60.5** |
| | Intensive green roof/Semi-intensive green roof/Extensive green roof | 11.9 | | Natural pollinator`s modules | 21.0 | | Planting and renewal urban trees | 18.6 |
| | Smart roof | 11.1 | | Compacted pollinator`s modules | 7.6 | | Boulevards | 8.5 |
| | Others | 13.5 | | Others | 17.6 | | Others | 12.4 |
| **NBS1** | **Green corridors** | **60.7** | **NBS ¿** | **Large urban park** | **44.3** | **NBS ¿** | **Pocket garden/park** | **58.6** |

| | | | |
|---|---|---|---|
| | Green corridors and belts | 36.7 | |
| | Others | 2.7 | |

| NBS20 | | | |
|---|---|---|---|
| | **Large urban public park** | **25.0** | |
| | Green resting areas | 13.6 | |
| | Others | 17.1 | |

| NBS21 | | | |
|---|---|---|---|
| | Green resting areas | 20.7 | |
| | Parklets | 10.7 | |
| | Others | 10.0 | |

| NBS19 | | |
|---|---|---|
| | **Urban forest** | **85.1** |
| | Others | 14.9 |

| NBS20 | | |
|---|---|---|
| | **Heritage garden** | **67.6** |
| | Heritage park | 23 |
| | Others | 9.4 |

| NBS21 | | |
|---|---|---|
| | **Private gardens** | **97.9** |
| | Others | 2.1 |

| NBS22 | | |
|---|---|---|
| | **Community garden** | **55.6** |
| | Vegetable gardens | 30.3 |
| | Others | 14.1 |

| NBS23 | | |
|---|---|---|
| | **Urban Orchard** | **68.3** |
| | Community garden | 13 |
| | Others | 18.7 |

| NBS25 | | |
|---|---|---|
| | **Composting** | **72.9** |
| | Community composting | 25 |
| | Others | 2.1 |

| NBS24 | | |
|---|---|---|
| | **Use of pre-existing vegetation** | **55.7** |
| | Protected areas | 6.8 |
| | Ecosystem conservation | 3.4 |
| | Others | 34.1 |

| NBS26 | | |
|---|---|---|
| | **Soil improvement and conservation measures** | **36.7** |
| | **Soil improvement** | **31.7** |
| | Mulching | 7.5 |
| | Enhanced nutrient managing and releasing soil | 6.7 |

| NBS27 | | |
|---|---|---|
| | **Systems for erosion control** | **41.9** |
| | **Soil & slope revegetation** | **29.5** |
| | Strong slope vegetation | 7 |
| | Others | 21.7 |

| | | | | | | | | |
|---|---|---|---|---|---|---|---|---|
| | | | | Others | 17.5 | | | |

| | | | |
|---|---|---|---|
| Others | 20.4 | Permeable pavements | 11.2 |
| | | Others | 19.2 |

Table A.4 Results from one sample chi-square test based on P-value method with significance level of 0.05 (α).

| Variable | Name | % of valid votes | ¹Total valid votes | X² value | P-value |
|---|---|---|---|---|---|
| **NBS1** | *(Dry) Detention Pond* | 36.0 | 86 | 6.70 | 0.01 |
| | *Infiltration basin* | 64.0 | | | |
| **NBS4** | *Swale* | 43.9 | 66 | 0.97 | 0.32 |
| | *Bioswale* | 56.1 | | | |
| **NBS5** | *Constructed wetlands* | 51.7 | 87 | 0.10 | 0.75 |
| | *Constructed wetland for wastewater treatment* | 48.3 | | | |
| **NBS6** | *Climber green wall* | 40.7 | 81 | 2.78 | 0.10 |
| | *Green façade with climbing plants* | 59.3 | | | |
| **NBS17** | *Large urban public park* | 38.8 | 85 | 4.25 | 0.04 |
| | *Large urban park* | 61.2 | | | |
| **NBS26** | *Soil improvement* | 62.0 | 79 | 4.57 | 0.03 |
| | *Soil improvement and conservation measures* | 38.0 | | | |
| **NBS27** | *Soil & slope revegetation* | 27.2 | 81 | 16.90 | 3.94E-05 |
| | *Systems for erosion control* | 72.8 | | | |
| **NBS28** | *Systems for erosion control* | 47.4 | 78 | 0.21 | 0.65 |
| | *Vegetation engineering systems for riverbank erosion control* | 52.6 | | | |
| **NBS29** | *Rivers or streams, including re- meandering, re-opening Blue corridors* | 81.0 | 58 | 22.34 | 2.3E-06 |
| | *Daylighting* | 19.0 | | | |
| **NBS30** | *Floodplain* | 46.3 | 80 | 0.45 | 0.50 |
| | *Reprofiling/Extending floodplain area* | 53.8 | | | |
| **NBS31** | *Systems for erosion control* | 36.8 | 68 | 4.76 | 0.03 |
| | *Diverting and deflecting elements* | 63.2 | | | |
| **NBS32** | *Green parking pavements* | 38.5 | 78 | 4.15 | 0.04 |
| | *Vegetated grid pave* | 61.5 | | | |

[1]Total valid votes stands for total votes minus blanks and option "I don't know/I have no opinion".

Table A.5 Results of Scopus search performed in October 2020 (In bold the names adopted).

| NBS Nº | Possible names | Scopus search | Nº of documents |
|---|---|---|---|
| 4 | **[1]Swale** | **Swale*** | **2.068** |
|  | [1]Bioswale | Bioswale* | 135 |
| 5 | **[1]Constructed Wetland** | "Constructed wetland*" | **9.448** |
|  | [1]Constructed wetland for wastewater treatment | "Constructed wetland* for wastewater treatment" | 177 |
| 6 | [1]Green façade with climbing plants | "Green façade* with climbing plant*" OR "Green facade* with climbing plant*" | 0 |
|  | [1]Climber green wall | "Climb* green wall*" | 4 |
|  | [2]Soil-based green façade | "Soil-based green façade*" OR "Soil-based green facade*" OR "Soilbased green façade*" OR "Soilbased green facade*" OR "Soil based green façade*" OR "Soil based green facade*" | 0 |
|  | **[2]Green façades** | "green façade*" OR "green facade*" | **236** |
| 28 | [1]Vegetation engineering systems for riverbank erosion control | "Vegetation engine* systems for riverbank* erosion control" OR "Vegetation engine* systems for river bank* erosion control" | 0 |
|  | [1]Systems for erosion control | "System* for erosion control" | 7 |
|  | **[2]Riverbank engineering** | "Riverbank* engine*" OR " River bank* engine*" | **1** |
| 30 | [1]Floodplain | Floodplain* OR "Flood plain*" | 30.330 |
|  | **[1]Reprofiling/extending floodplain** | "Reprof* floodplain*" OR "Reprof* flood plain*" OR "extend* floodplain*" OR "extend* flood plain*" OR "floodplain* extension*" OR "flood plain* extension*" | **19** |

[1] Two most voted names from the first survey in which there was no consensus during the second survey. [2] names suggested by experts.

Table A.6 Integrative List of NBS containing the final name, other possible names (survey and existing standards) and brief description.

| Acronym | Name | Other suggested name<br>[1]Survey or [2]existing standards | Brief Description |
|---|---|---|---|
| NBS1 | *Infiltration basin* | [1](Dry) Detention Pond; Floodable park; Wadi; Non-permanent infiltration basin; Green water storage and infiltration system; Storm basin; Micro-catchment; The sponge zone.<br><br>[2] Wet meadows; Detention pond, dry pond (University of Arkansas Community Design Center, 2010). | This NBS is a shallow impound area with highly permeable soils designed to temporarily detain and infiltrate stormwater runoff. It collects and stores runoff, allowing it to infiltrate into the ground (allowing pollutants to settle and filter out). The water fills up the depression and then soaks into the ground or is discharged to a conveyance system or a receiving waterbody (in the latter case, the term detention pond can be used). This NBS does not retain a permanent pool of water. Adapted from URBANGREENUP (2018) (2018), UNALAB (2019) (2019), Woods-Ballard et al. (2015) and University of Arkansas Community Design Center (2010). |
| NBS2 | *(Wet) Retention Pond* | [1]Water Retention ponds; Green retention pond; Extended Retention Basin; Holding pond; Pond; Waste stabilization ponds with Bunds; (wet) retention basin<br><br>[2] Retention pond (Woods Ballard et al 2015); wet pool or wet pond (University of Arkansas Community Design Center, 2010) | This NBS consists of a constructed stormwater pond that retains a permanent pool of water and thus, provides additional storage capacity during rainfall events. It has the capacity to retain stormwater continuously, remove urban pollutants and release the effluent at a controlled rate. During dry periods it also holds water. Adapted from URBANGREENUP (2018), UNALAB (2019) and University of Arkansas Community Design Center (2010). |
| NBS3 | *Rain garden* | [1]Infiltration garden; Rainfall garden; Water control garden, Floodable garden, Bio retention filter, Bio retention area, Bioremediation wet retention. | It is a planted depression designed to collect, store, infiltrate and filter stormwater runoff on a small-scale, especially in urban areas. It combines layers of organic sandy soil for infiltration and mulch to promote microbial activity. Native plants are recommended based upon their intrinsic synergies with local climate, soil, and moisture conditions without the use of fertilizers and chemicals. Storm water runoff is drained, stored for a certain period, |
| | | [2]Bioretention facility (University of Arkansas Community Design Center, 2010). | and then infiltrates into the ground soil. This NBS can have an above-ground overflow for excess water, although in some instances a simple underdrain may be more effective than providing a small control structure or overflows. Adapted from URBANGREENUP (2018), UNALAB (2019), University of Arkansas Community Design Center (2010) and Woods-Ballard et al. (2015) |
| NBS4 | *Swale* | [1]Grassed swale; Green drainage corridor; Vegetative filter; Vegetated Bioswale;<br><br>[2]Bioswale (University of Arkansas Community Design Center, 2010) | This NBS is an open lined or unlined, gently sloped, vegetated channel designed for treatment and conveyance of stormwater runoff. The main function of a Swale is to treat stormwater runoff as it is conveyed, whereas the main function of a rain garden is to treat stormwater runoff as it is infiltrated. Bioswales require curb cuts, gutters or other devices that direct flow towards them. This NBS is often used to drain roads, paths or car parks while enhancing access corridors or other open spaces. Adapted from URBANGREENUP (2018), NATURE4CITIES (2020), UNALAB (2019), Woods Ballard et al 2015 and University of Arkansas Community Design Center (2010) |
| NBS5 | *Constructed wetland* | [1]Planted horizontal/vertical filter Helophyte filter; Root-zone Wastewater Treatment; Natural wastewater treatment<br><br>[1,2]Treatment wetlands; Artificial Wetland; Subsurface constructed wetland; Planted sand/soil filters; (Kadlec and Wallace, 2009). | This NBS includes a range of engineered systems designed and constructed to replicate natural processes occurring in natural wetlands involving vegetation, soils, and the associated microbial assemblages to assist in treating wastewater streams (e.g., domestic, industrial) and stormwater. The treated water can be reused for safe aquifer recharge or for other non-potable reuses such as irrigation and toilet flushing. This NBS can be divided in two main hydrological categories: **Free water surface wetlands**, a shallow sealed basin or sequence of basins (open water areas) containing floating plants, submerged plants or emergent plants (similar in appearance to natural marshes); **Subsurface flow wetlands,** which include Horizontal flow (HF) wetlands and Vertical flow (VF) wetlands. In this case, the water flows beneath the surface level, either horizontally or vertically, through the filter bed. Adapted from URBANGREENUP (2018), NATURE4CITIES (2020), UNALAB (2019), Somarakis et al. (2019), Kadlec and Wallace (2009), Vymazal (2010), University of Arkansas Community Design Center(2010) and Dotro et al. (2017). |

| NBS6 | *Green facade* | [1]Green facade with climbing plants; Climber green wall; Ground-based green-wall; Green climber wall; Green wall with ground-based greening; Climber plant wall; Ground-Based Green Facade with Climbing Plants; [2]Soil-based green facade | This NBS is based on the application of climbing plants along the wall (in building facade or other types of walls). The wall is completely or partially covered with greenery and the plants can grow upwards. The climber plants are planted in the ground (soil) or in containers (filled with soil or substrate) and grow directly on the wall (direct systems), or climb using climbing-aids (indirect systems) that are attached to the wall. Adapted from URBANGREENUP (2018), UNALAB (2019), NATURE4CITIES (2020) and Manso and Castro-Gomes (2015). |
| NBS7 | *Green wall system* | [1]Hydroponic green facade; Facade-bound greening; Facade bound green wall; living wall; Continuous green wall; Plant wall system; Green façade with vertical panels; Greening vertical panel; Vertical greening panel; [2]Wall-based green facade | This NBS is a greening vertical panel that is fixed onto walls (in building facade or other types of walls). The panels allow the placement of plants and substrate on the entire surface. In contrast to green façades, this NBS can support a wider variety of plant species (e.g. shrubs, grasses and several perennials). The panels can be continuous or modular. Continuous panels are lightweight and permeable screens in which plants are inserted individually. A frame is fixed to the wall to support the panel, normally leaving a gap between the system and the surface. Modular panels can have several elements with specific dimension (e.g.trays, vessels, planter tiles flexible bags), which include the growing media where plants can grow. These elements together can create modular sections that are supported by a complementary structure or fixed directly on the vertical surface. Some systems allow the removal of panels during wintertime. Adapted from URBANGREENUP (2018), UNALAB (2019), NATURE4CITIES (2020), Manso and Castro-Gomes (2015). |

| NBS8 | *Vertical mobile garden* | [1]Mobile vertical greening; Mobile Green Living Room; Mobile green wall; Mobile vertical garden; Portable Green Wall; Mobile planter | This NBS is a vertical, mobile, planted, self-supporting module. It is fixed to a hook lift container platform. On this structure, different layers are placed along a hydroponic substrate in which the plants can grow. This NBS can be located anywhere in the city. Adapted from URBANGREENUP (2018) and UNALAB (2019). |
|---|---|---|---|
| NBS9 | *Planter green wall* | [1]Planters; Planter green facade; Planter boxes; Planter pots; Planter-based green wall; Planted/planting container(s); Pot planted plants; Potted plants; Potted Mobile Garden; Raised bed; container plants [2]Pot-based green facade | This NBS involves the use of planted containers such as pots or planters, filled with artificial (technical) soilless substrate or soil or a mixture. They can be placed individually or grouped on the ground or directly on the building façade or balconies. They can be used with almost any kind of plants, e.g. climbing plants, trees and/or shrubs. Adapted from NATURE4CITIES (2020). |
| NBS10 | *Vegetated pergola* | [1]Green pergola; Greened Pergola; Green matrasses Green shady structures; green shade | This NBS uses pillars, beams, stretched textile structure and lattices in different materials and compositions to create a growing assistance for vegetation and provide shaded areas. On this structure an inert substrate can be installed, to be covered with seeds. This NBS can be fixed to the facades of the buildings, on the street or by posts fixed to the sidewalk. Adapted from URBANGREENUP (2018) and NATURE4CITIES (2020). |
| NBS11 | *Extensive green roof* | [1]Green roof; Constructed wet roof; Green covering shelters; living roof; vegetated roof | This NBS involves basic, lightweight, planted systems that are implemented on the rooftop of building and are usually not accessible by the public. The most common plants used are sedum, herbs, mosses, and grasses. The installation and maintenance are less expensive than that of intensive systems. The growth medium is relatively thinner (8-15 cm) than for intensive systems (more than 20 cm). Adapted from URBANGREENUP (2018), UNALAB (2019), NATURE4CITIES (2020), Somarakis et al. (2019) and FFL (2002) |
| NBS12 | *Intensive green roof* | [1]Intensive/semi-intensive green roof; Roof garden; Roof park; Public Intensive Green Roof; Social Intensive Green Roof; Vegetated roof; Living roof | This NBS consists of implementing a great diversity of vegetation (higher variety than extensive GR) on rooftops. These spaces are normally accessible to the public for recreation, gardening, relaxation and socializing purposes. This NBS is usually heavier and has a deeper substrate (more than 20 cm) as compared to extensive systems. In addition, it requires higher installation and maintenance efforts such as regular irrigation and fertilisation but |

| | | | provides more biotopes and higher biodiversity. Adapted from URBANGREENUP (2018), UNALAB (2019), NATURE4CITIES (2020) and FFL (2002). |
|---|---|---|---|
| NBS13 | *Semi-intensive green roof* | [1]Smart roof; Biodiversity roof; Eco systemic roof | This NBS is implemented on rooftops as a combination of areas of intensive and extensive green roofs. This NBS can support a great variety of plants such as small herbaceous plants, ground covers, grasses, perennials and small shrubs, as well as higher growing plants). Plants with lower water and fertilizer demand are prioritized, requiring moderate maintenance (more than extensive and less than intensive systems). Adapted from NATURE4CITIES (2020), UNALAB (2019) and FFL (2002). |
| NBS14 | *Create and preserve habitats and shelters for biodiversity* | [1]Biodiversity micro habitats; Biodiversity facilitating structures; shelter for biodiversity; Shelters for insects /shelters for animals bird modules/insect modules; Habitat elements; Habitat replicate; Habitat features; Habitat enhancement; Habitat structures; Habitat and shelter modules; Artificial shelters ('tree hollows', 'pollinator planting', 'understorey planting'); BioDiv device. | This NBS replicates habitats for biodiversity and thus is designed to attract, shelter or provide food to a specific type of organism and animals or animal. It may involve creating or preserving existing site(s) or individual or grouped elements. Some examples of this NBS are: dead wood; standing dead trees; pollinator areas or modules; spot planted with vegetation of different age, size, type and features; trees displaying hollows or cavities; artificial shelters. Adapted from URBANGREENUP (2018) and NATURE4CITIES (2020). |
| NBS15 | *Street trees* | [1]Planting and renewing urban trees; boulevards; Allée; urban trees; urban tree canopy; Tree infrastructure; Urban trees alignment; single line trees; Sustainable management of urban trees; single tree<br>[2]Trees on streets or in public squares (FAO, 2016) | This NBS is focused on planting, renewing or maintaining urban street trees. It is designed to be appropriate for its context (right tree in the right place) and to achieve multiple benefits. One single or several trees can be arranged along streets, bicycle paths and sidewalks. These trees are situated on a single side (e.g. single line trees) and if circumstances allow, they can be established on both sides of the route (e.g.boulevard). In the latter case, the treetops of opposite trees often form a (nearly) closed canopy. Adapted from URBANGREENUP (2018), NATURE4CITIES (2020), UNALAB (2019) and FAO (2016). |

| NBS16 | *Green corridors* | [1]Green belts; Green way | This NBS aims to renature areas of derelict infrastructure such railway lines or along waterways and rivers, transforming them into linear parks. This NBS can be considered as a transitional area between biomes that connect neighbourhoods. This NBS can play an important role in urban green infrastructure networks. Adapted from NATURE4CITIES (2020) and Somarakis et al. (2019). |
|---|---|---|---|
| NBS17 | *Large urban park* | [1]Large urban public park; Urban park; Public park; Park; Green Park; Residential Park; Greened recreation areas/regional parks; Green resting areas [2]City park (FAO, 2016) | This NBS refers to large green areas (>0.5 ha) within a city with a variety of active and passive recreational facilities that meet the recreational and social needs of the residents and of visitors to the city. They are open to wide-range communities. This NBS can serve all the city or part of city, and it is open to wide-range communities. Adapted from URBANGREENUP (2018), NATURE4CITIES (2020), UNALAB (2019) and FAO (2016). |
| NBS18 | *Pocket garden/park* | [1]Small Park; Neighbourhood park; Landscape park; Empowerment Park [2]Pocket parks (FAO, 2016) | This NBS includes publicly accessible and compact green areas or small gardens (<0.5 ha) around and between buildings vegetated by ornamental trees, grass and other types of plants. It is projected for resting, relaxation, observing nature and social contact. This NBS provides opportunities for people to create small but important public spaces right in their own neighbourhoods. Adapted from URBANGREENUP (2018), NATURE4CITIES (2020) and FAO, (2016). |
| NBS19 | *Urban forest* | [1]Group of trees; Wood; Urban woodland; Arboreal areas around urban areas; Arboreal urban parks; Arboretum; Urban tree cover | This NBS mimics the appearance/form of a forest in an urban setting. It comprises all woodlands, groups of trees, and individual trees, forests, street trees, trees in parks and gardens, and trees in derelict corners. Adapted from URBANGREENUP (2018), UNALAB (2019), NATURE4CITIES (2020) and FAO (2016). |
| NBS20 | *Heritage garden* | [1]Heritage park/garden; Historical park/garden | This NBS refers to historical, long-appreciated gardens which preserve outstanding social, cultural, aesthetic or scientific value. Adapted from NATURE4CITIES (2020). |
| NBS21 | *Private gardens* | [1]Residential garden; House garden | This NBS is an urban green area, privately owned or rented, in the immediate vicinity of private houses. It is cultivated mainly for ornamental purposes and/or non-commercial food |

| | | | production. It offers possibilities to sustain and improve ecological and social qualities in people's everyday environment. Adapted from NATURE4CITIES (2020). |
|---|---|---|---|
| NBS22 | *Community garden* | [1]Allotment(s); community allotment(s); urban allotment; Urban garden/gardening; Urban agriculture; Urban farm; Multi cultivated gardens; Edible city solution; Semi- subsistence garden; Food growing area; Community Growing Spaces; P-Patches | This NBS is an area of land dedicated to the cultivation of vegetables, fruits and flowers for the purpose of food production. This kind of solution is usually encountered in public spaces, allotments or private residential properties. Adapted from NATURE4CITIES (2020). |
| NBS23 | *Urban Orchard* | [1]Community orchard; Community urban orchard; Allotment orchard; Food forest | This NBS is a piece of land in an urban area dedicated to the cultivation arboreal species. It is considered a social/community space. In general, the cultivation is based on organic premises and non-profit associations, neighbourhood associations or the city council are the most common management entities. Adapted from URBANGREENUP (2018) and NATURE4CITIES (2020). |
| NBS24 | *Use of pre-existing vegetation* | [1]Conservation of existing habitats; Use of native vegetation; Protected previous ecosystems; Protected areas; Conservation of existing habitats; Ecosystem conservation; Keep existing habitat; Indigenous gardening. | This NBS stands for the action of preserving a total or part of pre-existing ecosystems and vegetation in a certain area in order to protect essential habitats and to reduce implementation/maintenance costs of green areas. Adapted from NATURE4CITIES (2020). |
| NBS25 | *Composting* | [1]Community composting; Compost heap; Composting facility | This NBS includes all the structures and procedures required to compost food waste, vegetable materials, waste from cleaning grain, crop residues, etc. Adapted from URBANGREENUP (2018) and NATURE4CITIES (2020). |

| NBS26 | *Soil improvement* | [1]Soil improvement and conservation measures; Soil enhancement(s); Gentle remediation options; Soil management; Engineered, improved soil; Soil amendment | This NBS includes several techniques and methods to maintain and enhance soil quality in terms of physical, chemical and biological features. It aims to improve nutrient management, increase carbon storage, enhance water infiltration, encourage beneficial soil organisms and prevent soil compaction. Some techniques related to this NBS are: application of biochar, mulching, use of leguminous species for enhancing nitrogen fixation, use of organic matter, retaining stubble and green manuring to increase organic content and reduce compaction and erosion, organic fertilizer that stimulate and increase the soil biological activity and diversity. Adapted from URBANGREENUP (2018) and NATURE4CITIES (2020). |
| --- | --- | --- | --- |
| NBS27 | *Systems for erosion control* | [1]Soil-bioengineering (slope); Soil (and Water) bioengineering for slope stabilization and erosion control; Soil & slope revegetation; Strong slope vegetation; slope vegetation/revegetation; Slope stabilisation through revegetation; Soil and slope stabilisation; Vegetation engineering systems for slope erosion control | This NBS includes a set of different soil bioengineering techniques to stabilise soil structure on steepened slopes, to minimise/prevent the erosion of soil from wind or water, landslides and sedimentation problems. Common techniques are: revegetation (plants with strong deep roots), hydro-seeding, erosion control mat, covering natural fibre mats, wooden structures, and surface roughening. Adapted from NATURE4CITIES (2020). |
| NBS28 | Riverbank engineering | [1]Vegetation engineering systems for riverbank erosion control; Bioengineering (soil, water, fluvial, riverbanks); Riverbank stabilization/slope stabilization; Vegetated bank protection; Systems for erosion control on riverbanks; Riverbank protection system | This NBS includes techniques used in fluvial bioengineering for riverbank protection and hillside stabilisation to reduce risk of erosion by generating a natural protection. Some techniques embraced by this NBS are: Planted embankment mat; Plants are established on hills with strong inclination to provide strong and branched root networks; engineered designs using plant material and woody plant parts (e.g. fascine constructions; willow branch mattress); Living and dead wood can be combined (e.g. Vegetated crib walls, dead and live wood branch packing) for linear application and wide-spread effects; live stakes and other plant elements can be used jointly or individually, to stabilise the slope (live stakes, root stocks, fascine brushes etc.). Adapted from UNALAB (2019) and NATURE4CITIES (2020). |

| NBS29 | *Rivers or streams, including re-meandering, re-opening Blue corridors* | [1]Soil- bioengineering for River Re-naturing; River restoration; River revitalization; Daylighting; Reopened stream; Channel widening and length extension; Reprofiling the channel cross-section; Channel reprofiling and re-opening; Fluvial restoration/rehabilitation; Deculverting and re-meandering | This NBS includes a set of techniques used in fluvial bioengineering to increase channel storage capacity and optimize flow dynamics in order to reduce flood risk and erosion. Some examples are: the river channel is widened or deepened to enhance the flood dissipation capacity; the channel of covered/buried watercourses is opened by removing concrete layers. Adapted from URBANGREENUP (2018), NATURE4CITIES (2020) and UNALAB (2019). |
| --- | --- | --- | --- |
| NBS30 | *Reprofiling/extending floodplain* | [1]Floodplain; Branches; Floodplain widening; Floodplain restoration; restore /increase the floodplain area; Room-for-the-river approach / Floodplain management; Extending flood plain area; Management of the fluvial space River morphology and floodplain restoration; Floodplain excavation/enlargement/restoration | This NBS includes a set of techniques used in fluvial bioengineering reprofiling or extending the floodplain in order to provide additional flooding space and thus, reduce flood risk. During low water levels, these relatively flat and accessible bank areas can be used for recreational or agricultural purposes. This NBS includes the following techniques: Excavating the lateral riverbed; Dividing the discharge into two branches (by-passes, creating islands). Adapted from NATURE4CITIES (2020) and UNALAB (2019). |
| NBS31 | *Diverting and deflecting elements* | [1] Instream structures; Natural flow diversion structures; (Soil and) Water Bioengineering for stream restoration; Water bioengineering flow changing techniques; Redirection of the water flow, Stimulation of river dynamic processes; Riverbed morphology engineering; Increased water course friction | This NBS employs elements such as rocks, larger tree trunks, willow branches that are placed near the riverbank or in the middle of a river. This NBS alters flow variation and sediment shifting processes affecting the development of the channel's length and depth. In this sense, the main objective is to redirect, disturb, divert and deflect the water flow and initiate water dynamics for riverside protection against erosion. Adapted from URBANGREENUP (2018) and UNALAB (2019). |

| NBS32 | *Vegetated grid pave* | [1] Permeable/pervious/infiltration pavements; Green/greened/vegetated/grass pavements; green parking pavements; Engineered Vegetated Green Pavement; grass block paver/interlocking grass paver; Permeable pavements and parking lots; Pervious surfacing; Permeable green pavements | This NBS includes planted pavement structures normally filled with soil, grass seeds, gravel or rocks. It can be considered as a type of pervious/permeable pavement. The runoff soaks through the pavement structure and can be stored or infiltrated into the ground. The structures are modular and adaptable to different surface types such as parking areas, roadways, cycle-pedestrian paths, sidewalks or street furniture zones. Usually, the costs and maintenance are low compared to traditional pavements. Adapted from URBANGREENUP (2018), NATURE4CITIES (2020), UNALAB (2019), Woods Ballard et al 2015 and University of Arkansas Community Design Center (2010). |

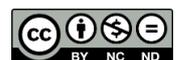